\documentclass[a4paper,11pt]{article}
\pdfoutput=1
\usepackage{amsmath,amssymb}
 \usepackage[utf8]{inputenc}
\usepackage{mathtools}
\usepackage{tikz-cd}
\usepackage{tikz}
\usetikzlibrary{arrows.meta}
\usetikzlibrary{arrows,calc,snakes,shapes,shapes.arrows,
decorations.markings,decorations.pathmorphing}
\usepackage{arydshln}
\usepackage[colorlinks=true,citecolor=blue,linkcolor=blue,urlcolor=blue]{hyperref}
\numberwithin{equation}{section}
\usepackage{xcolor}
\usepackage{geometry}
\def\makeatletter{\catcode`\@=11}
\makeatletter
\def\mathbox#1{\hbox{$\m@th#1$}}%
\def\math@ccstyles#1#2#3#4#5#6#7{{\leavevmode
      \setbox0\mathbox{#6#7}%
      \setbox2\mathbox{#4#5}%
      \dimen@ #3%
      \baselineskip\z@\lineskiplimit#1\lineskip\z@
      \vbox{\ialign{##\crcr
             \hfil \kern #2\box2 \hfil\crcr
             \noalign{\kern\dimen@}%
             \hfil\box0\hfil\crcr}}}}
\def\mathaccstyles{\math@ccstyles\maxdimen}
\def\maththroughstyles{\math@ccstyles{-\maxdimen}}
\def\unity%
 {\maththroughstyles{.45\ht0}\z@\displaystyle {\mathchar"006C}\displaystyle 1}

\makeatletter
\newcommand\longleftrightarrowfill@{%
  \arrowfill@\leftarrow\relbar\rightarrow}
\makeatother

\begin{document}

\begin{flushright}
\footnotesize
\texttt{DESY 19-045}

\texttt{Imperial/TP/2019/AB/01}
\vspace{2.1cm}
\end{flushright}

\centerline{\LARGE \bf Discrete gauge theories of charge conjugation}

\vspace{2truecm}

\centerline{
    {\large \bf Guillermo Arias-Tamargo, ${}^{a}$} \footnote{ariasguillermo@uniovi.es}
     {\large \bf Antoine Bourget, ${}^{b}$} \footnote{a.bourget@imperial.ac.uk}}
     \vspace{.3cm}
     \centerline{
    {\large \bf Alessandro Pini ${}^{c}$} \footnote{alessandro.pini@desy.de}
     {\bf and}
    {\large \bf Diego Rodr\'{\i}guez-G\'omez${}^{a}$} \footnote{d.rodriguez.gomez@uniovi.es}}

\vspace{1cm}
\centerline{{\it ${}^a$ Department of Physics, Universidad de Oviedo}} \centerline{{\it C.~Federico García Lorca 18, 33007, Oviedo, Spain}}
\vspace{.3cm}
\centerline{{\it ${}^b$ Theoretical Physics, The Blackett Laboratory, Imperial College}} \centerline{{\it
SW7 2AZ United Kingdom}}
\vspace{.3cm}
\centerline{{\it ${}^c$ DESY Theory Group}} \centerline{{\it Notkestraße 85, 22607 Hamburg, Germany}}
\vspace{2cm}

\centerline{\bf ABSTRACT}
\vspace{.5cm}

\noindent We define gauge theories whose gauge group includes charge conjugation as well as standard $\mathrm{SU}(N)$ transformations. When combined, these transformations form a novel type of group with a semidirect product structure. For $N$ even, we show that there are exactly two possible such groups which we dub $\widetilde{\mathrm{SU}}(N)_{\mathrm{I,II}}$. We construct the transformation rules for the fundamental and adjoint representations, allowing us to explicitly build four-dimensional $\mathcal{N}=2$ supersymmetric gauge theories based on $\widetilde{\mathrm{SU}}(N)_{\mathrm{I,II}}$ and understand from first principles their global symmetry. We compute the Haar measure on the groups, which allows us to quantitatively study the operator content in protected sectors by means of the superconformal index. In particular, we find that both types of $\widetilde{\mathrm{SU}}(N)_{\mathrm{I,II}}$ groups lead to non-freely generated Coulomb branches.

\newpage

\newpage
{
\hypersetup{linkcolor=black}
\tableofcontents
}

\section{Introduction}

Gauge symmetry governs the dynamics of a huge variety of systems, ranging form Condensed Matter to Particle Physics. Very often, when discussing gauge symmetry, one implicitly refers to symmetries associated to continuous groups. However, gauge theories based on discrete groups (\textit{a.k.a.} discrete gauge theories), while perhaps more exotic, are also very interesting. Indeed, in Condensed Matter (or lattice models), discrete gauge theories play a relevant role. For instance, in 2d, a web relating various well-known dualities was recently described in \cite{Karch:2019lnn} by including appropriate $\mathbb{Z}_2$ gaugings using previous results in \cite{Kapustin:2014dxa,Kapustin:2017jrc}.  Also in 3d discrete gauge theories play a relevant role. For example, the $\mathbb{Z}_2$ Ising model for one-half spins in a squared lattice is dual to a lattice $\mathbb{Z}_2$ gauge theory \cite{doi:10.1063/1.1665530}. Moreover, it admits a phase whose continuum limit is realized by the same doubled Chern-Simons theory appearing in the description of certain topological phases of electrons in \cite{2004AnPhy.310..428F}.

In turn, in High Energy Physics, discrete gauge theories also play a relevant role. In many cases, discrete global symmetries -- a prominent example being R-parity in supersymmetric (SUSY) models -- are needed to achieve phenomenologically viable scenarios. Yet, if only global symmetries, their constraints would be washed out by Quantum Gravity effects. This suggests, as first discussed in \cite{Krauss:1988zc}, that discrete symmetries must be gauged at a fundamental level. The subject was recently revived in \cite{Banks:2010zn}, where it was argued that in a consistent theory of Quantum Gravity such as String Theory all global symmetries, including discrete ones, are expected to be gauged. Indeed, String Theory quite often produces gauged discrete symmetries. For instance, in the presence of NS5 branes, there can be discrete $\mathbb{Z}_k$ gauge symmetries as discussed in \cite{Maldacena:2001ss}. These gauged discrete symmetries have also been quite extensively discussed in the context of String Phenomenology (see \textit{e.g.} \cite{BerasaluceGonzalez:2011wy}).  More recently, \cite{Hanany:2018vph} conjectured discrete $S_N$ gauge symmetries in 6d Conformal Field Theories (CFT's) needed in order to correctly reproduce their operator spectrum.

On a seemingly separate line, the traditional approach to Quantum Field Theory (QFT) is based on perturbation theory, specifically through the computation of correlation functions using Feynman diagrams. While this approach is very successful to compute observables such as scattering amplitudes, it misses a great deal of the beautiful subtleties of Quantum Field Theory. Indeed, by definition perturbation theory considers small fluctuations around the (trivial) vacuum, and hence it is essentially blind to the global structure of the group, which at most enters as a superselection rule. Nevertheless, interesting Physics may be hiding in the global structure of the group, despite being local Physics blind to it. A particular example is the case of $\mathrm{O}(N)$ theories, which can be regarded as the composition of a continuous gauge $\mathrm{SO}(N)$ symmetry and a discrete gauge $\mathbb{Z}_2$ symmetry. Thus, this is yet another context in which discrete gauge theories may appear, in this case as part of a larger and disconnected gauge group.

Also in the realm of High Energy Physics, the gauging of discrete symmetries has been argued to play a very relevant role in the construction of $\mathcal{N}=3$ SCFTs in 4d. The first examples of these were constructed in \cite{Garcia-Etxebarria:2015wns} starting with $\mathcal{N}=4$ SYM where the complexified Yang-Mills coupling is tuned to a self-S-dual point. At those points a subgroup $\Gamma\subset \mathrm{SL}(2,\,\mathbb{Z})$ mapping the theory to itself appears as an extra discrete global symmetry. As shown in \cite{Garcia-Etxebarria:2015wns,Argyres:2016yzz,Bourton:2018jwb,Argyres:2018wxu}, quotienting by a well-chosen combination of $\Gamma$ and a subgroup of the R-symmetry -- which amounts to gauging a discrete group -- precisely breaks the supersymmetry down to exactly $\mathcal{N}=3$. Hence, it is the discrete gauging which is breaking the supersymmetry down to $\mathcal{N}=3$.

In \cite{Argyres:2016yzz} this strategy was generalized to a systematic study of gaugings of discrete symmetries preserving at least $\mathcal{N}=2$ SUSY (mostly restricting to rank one theories). In this way a beautiful landscape of theories interrelated among them emerged. In the case of theories based on an $\mathrm{SU}(N)$ gauge group, a natural such discrete symmetry to quotient by is charge conjugation. As noted in \cite{Argyres:2016yzz}, gauging charge conjugation can be subtle and all supersymmetries can be broken. The ultimate reason for this  lies in the fact that charge conjugation is essentially akin to complex conjugation ($\sim \mathbb{Z}_2$), the outer automorphism of $\mathrm{SU}(N)$. It is then intuitive that the combined group incorporating $\mathrm{SU}(N)$ and charge conjugation transformations cannot simply be the direct product $\mathrm{SU}(N)\times \mathbb{Z}_2$ but rather the semidirect product $\mathrm{SU}(N)\rtimes \mathbb{Z}_2$.\footnote{Schematically, this can be easily seen by writing any element in the Cartesian product $\mathrm{SU}(N)\times \mathbb{Z}_2$ as $g(U,\,\gamma)$, where $\gamma$ is either the identity or complex conjugation ($\sim$ charge conjugation) and $U$ a $\mathrm{SU}(N)$ matrix. A short computation shows that the multiplication rule is that of a semidirect product (more details in section \eqref{matrixrealization}).} This conflicts with the two-step procedure of first considering a $\mathrm{SU}(N)$ gauge theory and then gauging its $\mathbb{Z}_2$ charge conjugation symmetry, which implicitly assumes a direct product structure.

In \cite{Bourget:2019phe} a fresh approach to the problem was taken, namely, first constructing a Lie group which incorporates both $\mathrm{SU}(N)$ transformations as well as charge conjugation, and then using it to build gauge theories. Such groups had been introduced in the mathematical literature under the name of \textit{principal extensions} (in this case of $\mathrm{SU}(N)$) in the past \cite{de1956groupes,wendt2001weyl}, and made a brief appearance in the Physics literature in \cite{Bachas:2000ik, Maldacena:2001xj, Stanciu:2001vw} in the context of branes wrapping group manifolds. A more related set-up is that discussed in \cite{Schwarz:1982ec}, where after symmetry breaking one ends up with a remainder discrete charge conjugation symmetry which can produce Alice strings. While this was mostly with an eye on the $U(1)$ case, the non-abelian version plays a relevant role as well in the process of constructing orientifold theories, as discussed in \cite{Harvey:2007ab}. These groups have two disconnected components and are very similar to the orthogonal gauge theories briefly alluded above. Indeed, $\mathrm{O}(2N)$ is a principal extension of $\mathrm{SO}(2N)$ \cite{Bourget:2017tmt} (since $\mathrm{O}(2N+1)=\mathrm{SO}(2N+1)\times \mathbb{Z}_2$, this case is much more tractable and less interesting). Surprisingly, as shown in \cite{Bourget:2019phe} and soon after in \cite{Argyres:2018wxu,Bourton:2018jwb} for other discrete gaugings of the like, it turns out that the gauge theories based on principal extension of $\mathrm{SU}(N)$ provide the first examples of four-dimensional $\mathcal{N}=2$ theories with non-freely generated Coulomb branches. While \textit{a priori} no argument forbids theories with non-freely generated Coulomb branches (and indeed their putative properties had been studied \cite{Argyres:2017tmj}), in view of the lack of explicit examples it was widely believed that such theories would not exist. The theories proposed in \cite{Bourget:2019phe,Argyres:2018wxu,Bourton:2018jwb} then provide the first counterexamples to that conjecture.

In \cite{Bourget:2019phe} $\mathcal{N}=2$ SUSY theories where considered as proof-of-concept for gauge theories based on principal extensions, with the bonus that the first theories with non-freely generated Coulomb branches were discovered. Yet in principle one may construct gauge theories in arbitrary dimensions with any SUSY (including no SUSY) based on principal extension groups. In this paper we re-consider with more detail the construction of such groups. It turns out that the correct way to think about these groups is as extensions of $\mathrm{SU}(N)$ by its (outer) automorphism group (recall, $\sim \mathbb{Z}_2$). A detailed analysis shows that actually there are exactly two possible extensions, corresponding to two possible disconnected Lie groups which we dub $\widetilde{\mathrm{SU}}(N)_{\mathrm{I,II}}$. In a precise way that we describe below, these two types correspond to the classification of symmetric spaces of type A. To our knowledge, the existence of these two possible extensions and their construction has not appeared before. We then go on and explicitly construct gauge theories based on them, concentrating, like in \cite{Bourget:2019phe} as a proof-of-concept, on $\mathcal{N}=2$ SQCD-like theories. In particular, we analyze certain protected sectors of the operator spectrum using different limits of the Superconformal Index (SCI). This allows us to show that the corresponding Coulomb branches are isomorphic as complex algebraic varieties and both are not freely generated. As a by-product of the explicit construction of the Lagrangian of the theories, we can understand from first principles the global symmetry pattern emerging from the Higgs branch Hilbert series computation in \cite{Bourget:2019phe}.

The organization of this paper is as follows: we start in section \eqref{groups} describing the principal extensions of $\mathrm{SU}(N)$ as extensions of the $\mathbb{Z}_2$ outer automorphism group of $\mathrm{SU}(N)$ by $\mathrm{SU}(N)$. As anticipated, we find exactly two such possibilities which are in one-to-one correspondence with the Cartan classification of symmetric spaces of type $A$. In section \eqref{reps} we study aspects of the representation theory of these groups, paying special attention to the fundamental and the adjoint representations as well as to some of the invariants which can be formed out of them. We also construct the Weyl integration formula over the $\widetilde{\mathrm{SU}}(N)_{\mathrm{I,II}}$ groups. In section \eqref{gauge} we turn to Physics and construct $\mathcal{N}=2$ SQCD gauge theories based on $\widetilde{\mathrm{SU}}(N)_{\mathrm{I,II}}$. As a by-product, we will provide an \textit{a priori} understanding of the global symmetry groups of the resulting theories. In section \eqref{spectrum} we turn to the quantitative analysis of the theories using various limits of the superconformal index as diagnostics tool. To that matter, the integration formula previously developed in section \eqref{reps} plays a very relevant role. In particular, we will find that the $\widetilde{\mathrm{SU}}(N)_{\mathrm{I,II}}$ theories have non-freely generated Coulomb branches. Finally, we conclude in \eqref{conclusions} with some final remarks and open lines. For the interest of the reader, we postpone to the appendices several technical details.

\section{Construction of two disconnected groups}\label{groups}

The groups we are interested in are extensions of $\mathbb{Z}_2$ by a Lie group, which in this paper we will take to be $\mathrm{SU}(N)$. In \cite{Bourget:2019phe}, the so-called \emph{principal extension} were considered, but it turns out that although the group of outer automorphisms of $\mathrm{SU}(N)$ is $\mathrm{Out}(\mathrm{SU}(N)) \cong \mathbb{Z}_2$, there are in some cases more than one (in fact, exactly two) inequivalent ways of constructing a semi-direct product of $\mathrm{SU}(N)$ by $\mathrm{Out}(\mathrm{SU}(N))$. This section aims at studying this issue in depth.

For concreteness, we will start with a pedestrian approach to the problem, and we will see that the two disconnected groups arise in a natural way. Then we will provide a more abstract, but also more rigorous construction, of the semi-direct products. As we will explain, they are built from \emph{involutive} outer automorphisms (IOA) of $\mathrm{SU}(N)$, i.e. automorphisms which are their own inverse. In turn, we will see that these are classified by real forms of the complex Lie algebra $\mathfrak{sl}(N,\mathbb{C})$, or equivalently by symmetric spaces. This last feature will also help us understand the global symmetry of the gauge theories constructed in later sections.

\subsection{Explicit matrix realization}\label{matrixrealization}

Let's construct the (disconnected) gauge group of an $\mathrm{SU}(N)$ theory in which charge conjugation is gauged as well. In such a theory, the lowest-dimensional non-trivial representation has dimension $2N$, so we will construct our group as a $2N \times 2N$ matrix group. It has a subgroup, denoted $G$, which is isomorphic to $\mathrm{SU}(N)$ in the fundamental plus antifundamental representation:
\begin{equation}
\label{defG}
    G = \left\{ \mathbf{U} = \left(\begin{array}{cc} M & 0 \\ 0  &  M^{\star}\end{array}\right) \,\middle\vert\, M \in \mathrm{SU}(N) \right\} \cong \mathrm{SU}(N) \, ,
\end{equation}
where the star denotes complex conjugation. The charge conjugation is a $\mathbb{Z}_2$ group which exchanges the fundamental and antifundamental of $\mathrm{SU}(N)$, so it has to be of the form
\begin{equation}
    \Gamma_A = \left\{ \left(\begin{array}{cc} \unity & 0 \\ 0 &  \unity\end{array}\right) , \left(\begin{array}{cc} 0 & A \\ A^{-1} &  0\end{array}\right) \right\} \cong \mathbb{Z}_2 \, ,
\end{equation}
where $A \in \mathrm{SU}(N)$ is a matrix on which we will come back later. The total gauge group, which we call $\widetilde{G}_A$, is the image of the Cartesian product $G \times \Gamma_A$ under the multiplication map,
\begin{equation}
\label{defGtilde}
    \widetilde{G}_A = \left\{ g \gamma \mid g \in G , \gamma \in \Gamma_A \right\} \, .
\end{equation}
We have added the subscript $A$ to insist on the fact that this depends on the matrix $A$ chosen above. The product in this group is simply matrix multiplication. Thus for two elements $g \gamma$ and $g' \gamma '$ of $\widetilde{G}_A$, we have
\begin{equation}
    g \gamma \cdot g' \gamma ' = g \gamma   g' \gamma ' =  (g \varphi_{\gamma} (g') ) (\gamma  \gamma ') \, ,
\end{equation}
where the last rewriting is necessary for the product to be manifestly in the form $g'' \gamma ''$.
This is a semi-direct product structure, with
\begin{equation}
\begin{array}{cccc}
\varphi_{\gamma}: & G &\rightarrow & G \\
 & \mathbf{U} & \mapsto & \gamma\,\mathbf{U}\,\gamma^{-1}
 \end{array}
\end{equation}
Note that for the non-trivial element $\gamma \in \Gamma_A$ one has
\begin{equation}
\label{S}
\varphi_{\gamma}(\mathbf{U})=\left(\begin{array}{cc} A\,M^{\star}\,A^{-1} & 0 \\ 0 & A^{-1}\,M\,A\end{array}\right)\, .
\end{equation}
Writing down the requirement that this matrix belongs to $G$ defined in (\ref{defG}) leads to the condition that $A\,A^{\dagger}=\unity$ together with $A^{-1}\,M\,A = (A\,M^{\star}\,A^{-1})^{\star}$ for all $M \in \mathrm{SU}(N)$. From the last condition it follows that $M\,(A\,A^{\star})=(A\,A^{\star})\,M$, which is solved for $A\,A^{\star}=\lambda \,\unity$. Since $A\in \mathrm{SU}(N)$, by multiplying on the left by $A^{\dagger}$, we could as well write $A^{\star}=\lambda\,A^{\dagger}$. Likewise, we could take the complex conjugate of the equation to write $A^{\star}\,A=\lambda^{\star}\,\unity$. Multiplying now on the right by $A^{\dagger}$ leads to $A^{\star}=\lambda^{\star}\,A^{\dagger}$, which requires $\lambda\in \mathbb{R}$. Then, since $A\in \mathrm{SU}(N)$, by taking the determinant, $|{\rm det}(A)|^2=\lambda^N=1$. Hence, we obtain
\begin{equation}
\label{Scases}
    A\,A^{\star}=
    \begin{cases}
    \pm \unity  & \textrm{for }N \textrm{ even} \\
    + \unity  & \textrm{for }N \textrm{ odd} \, .
    \end{cases}
\end{equation}
Thus, all in all, we have found a family of matrix groups given by \eqref{defGtilde} where
\begin{itemize}
    \item for odd $N$ one needs $A^T=A$. This defines a group that we call $\widetilde{\mathrm{SU}}(N)_{\mathrm{I}}$.
    \item for even $N$ we have two cases:
    \begin{itemize}
        \item $A=A^T$: this defines a group that we call $\widetilde{\mathrm{SU}}(N)_{\mathrm{I}}$ (the even $N$ version).
        \item $A=-A^T$: this defines a group that we call $\widetilde{\mathrm{SU}}(N)_{\mathrm{II}}$.
    \end{itemize}
\end{itemize}

While this gives an intuitive construction of two different groups $\widetilde{\mathrm{SU}}(N)_{\mathrm{I}}$ and $\widetilde{\mathrm{SU}}(N)_{\mathrm{II}}$, several questions are left unanswered: why did we choose to represent $\widetilde{G}_A$ in the specific form (\ref{defGtilde})? Is the symmetry property of the matrix $A$ enough to characterize entirely the groups $\widetilde{G}_A$? Do this construction really yield two non-isomorphic groups?

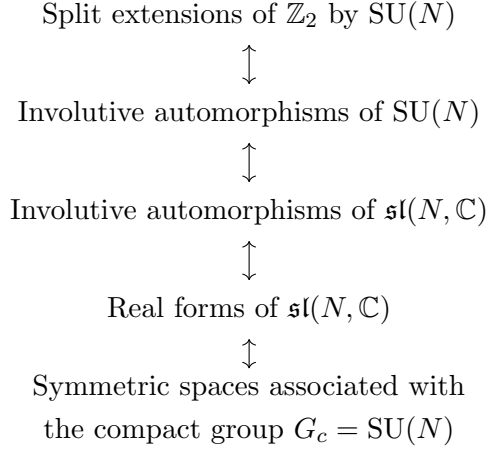
\begin{figure}[t]
\centering
\begin{tikzpicture}[decoration={markings,
mark=at position .5 with {\arrow{>}}}]
\begin{scope}[scale=1.3]
\node at (0,0) {Split extensions of $\mathbb{Z}_2$ by $\mathrm{SU}(N)$};
\node at (0,-1) {Involutive automorphisms of $\mathrm{SU}(N)$};
\node at (0,-2) {Involutive automorphisms of $\mathfrak{sl}(N,\mathbb{C})$};
\node at (0,-3) {Real forms of $\mathfrak{sl}(N,\mathbb{C})$};
\node at (0,-4) {\begin{tabular}{c}
     Symmetric spaces associated with  \\
     the compact group $G_c = \mathrm{SU}(N)$
\end{tabular}  };
\draw[<->=.5] (0,-0.7) to (0,-.3);
\draw[<->=.5] (0,-1.7) to (0,-1.3);
\draw[<->=.5] (0,-2.7) to (0,-2.3);
\draw[<->=.5] (0,-3.6) to (0,-3.3);
\end{scope}
\end{tikzpicture}
\label{figRoute}
\caption{Schematic representation of the method of classification of split extensions of $\mathbb{Z}_2$ by $\mathrm{SU}(N)$, covered in sections \ref{secExtensions} and \ref{secRealForms}.  }
\end{figure}

As for the last point, a preliminary observation is that had the two groups been conjugated one to the other, there should be an invertible $2N \times 2N$ matrix $\mathbf{X}$ such that $\gamma_{\mathrm{II}}=\mathbf{X}\,\gamma_{\mathrm{I}}\,\mathbf{X}^{-1}$, where we denoted by $\gamma_{\mathrm{I,II}}$ the non-trivial element of $\Gamma_A$ in the two cases. A natural ansatz for the matrix $\mathbf{X}$ is
\begin{equation}
    \mathbf{X}=\left(\begin{array}{cc} X & 0 \\ 0 & X^{\star}\end{array}\right)\,,
\end{equation}
for some $X\in \mathrm{SU}(N)$. A short computation shows that the condition for both choices to be conjugated translates into $A_{\mathrm{II}}=X\,A_{\mathrm{I}}\,X^T$, where $A_{\mathrm{I,II}}$ denotes in the obvious way the $A$ matrix for the corresponding choice. Transposing this equation leads to $A_{\mathrm{II}}=-X\,A_{\mathrm{I}}\,X^T$, which shows that such $X$ does not exist. While this hints that indeed both choices are two different groups, it strongly relies on a specific representation. In the rest of this section will offer more formal arguments that indeed there are exactly two extensions $\widetilde{\mathrm{SU}}(N)_{\mathrm{I}}$ and $\widetilde{\mathrm{SU}}(N)_{\mathrm{II}}$ of $\mathrm{SU}(N)$ for $N$ even, and only one for $N$ odd, that can be obtained by gauging the outer automorphism, and the labels $\mathrm{I}$ and $\mathrm{II}$ are related to the Cartan classification of real forms of $\mathfrak{sl}(N,\mathbb{C})$. Figure \ref{figRoute} gives an overview of the logical steps that we will follow.

\subsection{Extensions and semidirect products}
\label{secExtensions}

The semidirect products that we consider are extensions of $\mathbb{Z}_2$ by $\mathrm{SU}(N)$. In order to point out the subtleties of this construction, we begin with a review of some general theory of extensions of discrete groups by Lie groups (more details can be fount in \cite{hilgert2011structure}, Chapter 18). Let $\tilde{G}$ be any Lie group, connected or not. Then $\tilde{G}$ is the extension of the discrete groups of its connected components, called $\pi_0 (\tilde{G})$, by its identity component, called $G$. This means that there is a short exact sequence
\begin{equation}
\label{LieGroupExtension}
  \begin{tikzcd}
1 \arrow[r,] & G \arrow[r, "\iota"] & \tilde{G} \arrow[r, "q"]   & \pi_0(\tilde{G})  \arrow[r]   & 1  \,.
\end{tikzcd}
\end{equation}
Note that the groups introduced in the previous section fit this structure: constructing the maps $\iota(g)=g\,{\rm id}_{\Gamma_A}$ (which maps an element $g\in G$ into $\widetilde{g}\in\widetilde{G}$) and $q(\widetilde{g})=\gamma$ (which maps an element $\widetilde{g}=g\gamma \in\widetilde{G}$ into $\Gamma_A\cong \mathbb{Z}_2$), it is clear that ${\rm Ker}(q)={\rm Im}(\iota)$.

In the following we will restrict ourselves to \emph{split extensions}, or equivalently (due to the "splitting lemma") to the situation where $\tilde{G}$ is a semidirect product $G \rtimes \pi_0(\tilde{G})$. The assumption that the extension is split means that there exist a Lie group morphism $\sigma : \pi_0(\tilde{G}) \rightarrow \tilde{G}$ such that $q \circ \sigma = \mathrm{id}_{\pi_0(\tilde{G})}$. Note that for any extension (split or not) we can construct a map $C_G : \tilde{G} \rightarrow \mathrm{Aut}(G)$ defined by $C_G (\tilde{g}) (g) = \iota^{-1} (\tilde{g} \iota (g) \tilde{g}^{-1}) $ for all $g \in G$ and $\tilde{g} \in \tilde{G}$. Using now the splitting morphism $\sigma$ we can form a homomorphism $S = C_G \circ \sigma$, which is precisely what is needed to build a semi-direct product $G \rtimes_S \pi_0(\tilde{G})$.

Note as well that indeed the groups introduced in the previous section do fit in this structure, since we can construct a map from $\Gamma_A$ into $\widetilde{G}$ as $\sigma(\gamma)={\rm id}_G\,\gamma$ which clearly satisfies that $q\circ\sigma={\rm id}_{\Gamma_A}$. Hence, when regarded as a sequence, indeed $\widetilde{G}$ is split and, consequently, there is a semidirect product structure (which on the other hand we explicitly constructed). In that language, the homomorphism $S$ corresponds to the $\varphi_{\gamma}$ in eq. \eqref{S}.

Finally, we add a last ingredient to the construction, namely the group of outer automorphisms of $G$, $\mathrm{Out}(G) = \mathrm{Aut}(G) / \mathrm{Inn} (G)$, with the obvious map $[\cdot] : \mathrm{Aut}(G) \rightarrow \mathrm{Out}(G)$. For $\varpi \in \pi_0(\tilde{G})$, one can show that the formula $s(\varpi) = [C_G (\tilde{g})]$ where $\tilde{g} \in \tilde{G}$ is chosen such that $q(\tilde{g})=\varpi$ defined a group homomorphism $s : \pi_0(\tilde{G}) \rightarrow  \mathrm{Out}(G)$, called the \emph{characteristic homomorphism} of the extension. This is summarized by the diagram
\begin{equation}
  \begin{tikzcd}
1 \arrow[r,] & G \arrow[r, "\iota"] & \tilde{G} \arrow[r, bend left ,"q"]   \arrow[d, "C_G"]             & \pi_0(\tilde{G}) \arrow[l, ,"\sigma"]  \arrow[r]  \arrow[d, "s"]   \arrow[ld, "S"]  & 1 \\
  &                      & \mathrm{Aut}(G) \arrow[r,] & \mathrm{Out}(G)  &
\end{tikzcd}
\end{equation}
Two equivalent extensions of $\pi_0(\tilde{G})$ by $G$ define the same $s$.\footnote{See Lemma 18.1.6 in \cite{hilgert2011structure}. } Obviously, there are exactly two possible homomorphisms $s$  when $\pi_0(\tilde{G}) \cong \mathrm{Out}(G)  \cong  \mathbb{Z}_2$: the trivial morphism, giving the direct product $\tilde{G} = G \times \mathbb{Z}_2$, and the identity morphism, giving a semi-direct product. However the converse is not true, and we will see that two inequivalent extensions correspond to the identity morphism $s : \mathbb{Z}_2 \rightarrow \mathbb{Z}_2$. By definition, the classification of semidirect products $G \rtimes_S \pi_0(\tilde{G})$ reduce to the classification of the possible maps $S$. In the case at hand, where $\pi_0(\tilde{G}) \cong \mathbb{Z}_2$, $S$ has to be its own inverse, and thus this means that we need to classify the \emph{involutive} automorphisms of $G$. This is the main lesson that we learn from this abstract development: \emph{classifying the different extensions is equivalent to classifying the involutive automorphisms of $G$}, which will then be our next task. In particular, the semidirect products will be associated to outer involutive automorphisms.

\subsection{Real forms and Antiinvolutions}
\label{secRealForms}

There is a very elegant theory of involutive automorphisms in complex Lie algebras, connecting them to real forms -- this is what will allow us to classify them.
We begin with a quick reminder of some aspects of the theory of real and complex Lie algebras, referring to \cite{onishchik2004lectures} for more details. Given a real Lie algebra $\mathfrak{g}_{0}$ we denote by $\mathfrak{g}_{0}(\mathbb{C})$ the corresponding complexification, which is uniquely defined by the bracket
\begin{equation}
    [x_1+iy_1,x_2+iy_2] = [x_1,x_2] - [y_1,y_2] +i([x_1,y_2]+[y_1,x_2]), \  \forall \ x_1,x_2,y_1,y_2 \in \mathfrak{g}_0 \,.
\end{equation}
Conversely, a \textit{real structure} $\sigma$ of $\mathfrak{g}=\mathfrak{g}_{0}(\mathbb{C})$ is defined to be an involutive antilinear automorphism, or \emph{antiinvolution} for short. This means that a real structure is an automorphism which satisfies
\begin{equation}
\begin{cases}
     \sigma(\alpha x + \beta y ) = \alpha^{*}\sigma(x)+\beta^{*}\sigma(y) &  \forall \ x,y  \in  \mathfrak{g}_{0} \ \ \forall \alpha,\beta  \in  \mathbb{C}\\
     \sigma^2(z) = z & \forall \ z  \in  \mathfrak{g} \, ,
\end{cases}
\end{equation}
where $\alpha^{*}$ denotes the complex conjugate of $\alpha$.
Finally, a real subalgebra $\mathfrak{g}_0$ of $\mathfrak{g}$ is called a real form of $\mathfrak{g}$ if $\mathfrak{g}=\mathfrak{g}_{0} \oplus i\mathfrak{g}_{0}$.

It's important to note that although the complexification of a real Lie algebra is unique, there might be several real forms for a given complex Lie algebra, and these can be classified using the real structures. Indeed, on the complex Lie algebra $\mathfrak{g}$ there is a bijection between real structures and real forms:
\begin{itemize}
    \item Given a real structure $\sigma : \mathfrak{g} \rightarrow \mathfrak{g}$ we can construct the corresponding real form
    \begin{equation}
    \mathfrak{g}^{\sigma}:=\{X \in \mathfrak{g} \mid \sigma(X) = X \}\, .
    \end{equation}
    \item Conversely, given a real form $\mathfrak{g}_0$ of a complex Lie algebra $\mathfrak{g}$, we can construct the corresponding real structure $\sigma$ as the complex conjugation, $\sigma (x+iy) = x-iy$ \ $\forall$ \ $x,y \in \mathfrak{g}_0$.
\end{itemize}
Moreover two real forms $\mathfrak{g}_{0},\mathfrak{g}_{1}$ of $\mathfrak{g}$ are isomorphic if and only if the corresponding real structures $\sigma_0,\sigma_1$ are conjugate by an automorphism of $\mathfrak{g}$, i.e. there exists $\alpha \in \textrm{Aut} \ \mathfrak{g}$ such that $\sigma_1 = \alpha \sigma_0 \alpha^{-1}$.

\vspace{5mm}

We now focus on a rank $r$ semisimple complex Lie algebra $\mathfrak{g}$. It admits a canonical system of generators $(h_i , e_i , f_i)$ for $i=1 , \dots ,r$. One can show that \cite{onishchik2004lectures} there exists a unique real structure, which we call $\tau$ from now on, such that
\begin{equation}
    \tau (h_i) = -h_i \, , \qquad \tau (e_i) = -f_i \, , \qquad \tau (f_i) = -e_i \, .
\end{equation}
The associated real form is a \emph{compact} real form.\footnote{There exists also a unique real structure $\varsigma $ which fixes the system of generators,
\begin{equation}
    \varsigma (h_i) = h_i \, , \qquad \varsigma (e_i) = e_i \, , \qquad \varsigma (f_i) = f_i \, .
\end{equation}
The associated real form is the \emph{split} real form.  }

\begin{table}
\begin{equation*}
    \begin{array}{|c|c|c|c|} \hline
        \textrm{Cartan Class} & \textrm{Real form} \ \mathfrak{sl}(N,\mathbb{C})^{\sigma} & \textrm{Real structure } \sigma &  \textrm{Involution } \theta \\ \hline
         AI & \mathfrak{sl}(N,\mathbb{R}) & X \mapsto X^{\star} & \theta_{\mathrm{I}} : X \mapsto -X^T \\
         AII \textrm{ ($N$ even)} & \mathfrak{sl}(N/2,\mathbb{H}) &  X \mapsto - J_N X^{\star} J_N & \theta_{\mathrm{II}} : X \mapsto J_N X^T J_N \\ \hdashline
         AIII , AIV  & \mathfrak{su}(p,N-p)  & X \mapsto - I_{p,N-p} (X^{\star})^T I_{p,N-p} &  X \mapsto   I_{p,N-p} X I_{p,N-p} \\ \hline
    \end{array}
\end{equation*}
    \caption{The three (types of) real forms of the complex Lie algebra $\mathfrak{sl}(N,\mathbb{C})$. The second line exists only when $N$ is even. In the third line, $p=0,1,\dots , [N/2]$. For each real form, we indicate the corresponding real structure $\sigma$ and the corresponding involution $\theta = \sigma \tau$ with $\tau : X \mapsto - (X^{\star})^T$.  }
    \label{tableRealForms}
\end{table}

We now use the compact real structure $\tau$ to associate to any real structure $\sigma$ the automorphism
\begin{equation}
    \theta = \sigma \tau \,.
\end{equation}
It is clear that $\theta$ is a linear (as opposed to antilinear) automorphism, but in general it is not an involution. However, we have seen above that up to replacing the real form $\mathfrak{g}_{0}$ by an other isomorphic real form, we can conjugate the corresponding real structure $\sigma$ by any automorphism of $\mathfrak{g}$, and Cartan proved that at least one of these conjugates gives rise to an involutive $\theta$. Therefore, in the case of semisimple complex Lie algebras, to each real form one can associate an involutive automorphism. This non-trivial statement is the key step to obtain Cartan's theorem: two real forms $\mathfrak{g}_{0},\mathfrak{g}_{1}$ of $\mathfrak{g}$ are isomorphic if and only if the corresponding involutions $\theta_0 , \theta_1$ are conjugate to each other, i.e. $\exists \ \alpha \in \textrm{Aut}(\mathfrak{g}) \mid \theta_1 = \alpha \theta_0 \alpha^{-1}$.

With this theorem at hand, we can now, given a list of inequivalent real forms of a semisimple complex Lie algebra, obtain the corresponding classification of inequivalent involutive automorphisms. Let us work out the case of $\mathfrak{g} = \mathfrak{sl}(N,\mathbb{C})$. In that case, it is easy to check that\footnote{The split real form $\varsigma$ is the usual complex conjugation. } $\tau (X) = - (X^{\star})^T$ for $X \in \mathfrak{g}$. The real forms of $\mathfrak{sl}(N,\mathbb{C})$ can be read on Cartan's classification, and come in three different types; it is then a simple task to compute the associated involution $\theta$ in each case. This is summarized in Table \ref{tableRealForms}.
We have used the notations
\begin{equation}
    I_{p,q} =
    \left(
    \begin{array}{cc}
        -\unity_p & 0  \\
        0 & \unity_q
    \end{array}
    \right) \, , \qquad J_N =  \left(
    \begin{array}{cc}
        0 & -\unity_{N/2}  \\
        \unity_{N/2} & 0
    \end{array}
    \right) \, , \quad \textrm{($N$ even). }
\end{equation}
One representant of each conjugacy class of involution of $\mathfrak{sl}(N,\mathbb{C})$ is presented in the last column of Table \ref{tableRealForms}. Note that the involutions corresponding to the real forms $\mathfrak{su}(p,N-p)$ are inner, and we are left with precisely two conjugacy classes of outer involutive automorphisms when $N$ is even, and only one class when $N$ is odd.

There is a bijective correspondence between the simple real Lie algebras and the irreducible noncompact symmetric spaces of noncompact type which we briefly review in Appendix \ref{AppendixSymSpaces} (see \cite{bump2004lie}); this relates the involution of the algebra $\theta$ to  an involution $\Theta$ on the group. This correspondence is illustrated in the case of the compact group $\mathrm{SU}(N)$ in Table \ref{tableSymSpaces}. Although we will not exploit the symmetric spaces duality, we want to point out the involutions $\Theta$ and the subgroups $K$ left invariant by $\Theta$. One can check that in all cases the involutions $\Theta$ induce on the corresponding Lie algebras the involutions $\theta$ of Table \ref{tableRealForms}. As for the compact subgroups $K$, they will turn out to determine the global symmetry of gauge theories based on the semidirect products of $\mathrm{SU}(N)$ by $\Theta$.

From now on, we focus on the first two lines of Tables \ref{tableRealForms} and \ref{tableSymSpaces}, and borrowing names from the Cartan classification, we define the two following groups:
\begin{eqnarray}
\label{definitionSUtilde}
   && \widetilde{\mathrm{SU}}(N)_{\mathrm{I}} = \mathrm{SU}(N) \rtimes_{\Theta_{\mathrm{I}}} \mathbb{Z}_2\,,\nonumber \\ && \\ \nonumber
    &&\widetilde{\mathrm{SU}}(N)_{\mathrm{II}} = \mathrm{SU}(N) \rtimes_{\Theta_{\mathrm{II}}} \mathbb{Z}_2\,, \qquad \textrm{(} N \textrm{ even)} \,.
\end{eqnarray}
Note that these are indeed the groups constructed in the previous subsection, thus confirming the claim that indeed there are the two possible extensions of $\mathbb{Z}_2$ by $\mathrm{SU}(N)$.

\begin{table}
\begin{equation*}
    \begin{array}{|c|c|c|c|c|} \hline
        \textrm{Cartan Class} & G &  K  & \mathrm{dim } K &  \textrm{Involution } \Theta \\ \hline
         AI & \mathrm{SL}(N,\mathbb{R}) &   \mathrm{SO}(N)& \frac{1}{2}N(N-1) &  g \mapsto (g^{-1})^T \\
         AII \textrm{ ($N$ even)} & \mathrm{SL}(N/2,\mathbb{H}) &   \mathrm{Sp}(N/2) & \frac{1}{2}N(N+1) & g \mapsto -J_N  (g^{-1})^T J_N \\   \hdashline
         AIII , AIV  & \mathrm{SU}(p,N-p)  & S(\mathrm{U}(p) \times \mathrm{U}(N-p)) & p^2+(N-p)^2-1 & g \mapsto  I_{p,N-p} g I_{p,N-p} \\ \hline
    \end{array}
\end{equation*}
    \caption{The three (types of) symmetric spaces for which $G_c = \mathrm{SU}(N)$. In each case we indicate the dual group $G$, the compact subgroup $K$ and the lift to the group $\mathrm{SU}(N)$ of the involutions $\theta$ in Table \ref{tableRealForms}. One can check that $K$ is the subgroup of $G$ fixed by $\Theta$. }
    \label{tableSymSpaces}
\end{table}

\subsection{A construction of automorphisms}\label{sectionliftLiealg}

Now we explain how to construct explicitly automorphisms in the various classes corresponding to the lines of Table \ref{tableRealForms}. We will use a method based on the Weyl group.

\paragraph{General theory}
Consider a simple complex Lie algebra $\mathfrak{g}$.
Let $\phi : \Phi \rightarrow \Phi$ be an isomorphism of the root system $\Phi$, and let $\Delta$ be a set of simple roots in $\Phi$. The isomorphism $\phi$ extends in a trivial way on the Cartan subalgebra $\mathfrak{h}$, giving an isomorphism $\theta : \mathfrak{h} \rightarrow \mathfrak{h}$, and we want to extend it to the whole Lie algebra $\mathfrak{g}$. To do this, let us first choose a non-zero element $X_{\alpha}$ in each root space $\mathfrak{g}_{\alpha}$ for $\alpha \in \Delta$. We also choose a family of non-zero complex numbers $c_{\alpha}$ for $\alpha$ simple. Then (see \cite{humphreys2012introduction}, Theorem 14.2) there exist a unique isomorphism $\theta : \mathfrak{g} \rightarrow \mathfrak{g}$ that extends $\theta : \mathfrak{h} \rightarrow \mathfrak{h}$ and such that
\begin{equation}
\label{defcalpha}
    \theta (X_{\alpha} ) = c_{\alpha} X_{\phi (\alpha)}
\end{equation}
for every \emph{simple} root $\alpha \in \Delta$.

The Weyl group $W$, generated by reflections with respect to the hyperplanes orthogonal to the simple roots in $\mathfrak{h}^\ast$, corresponds to a set of automorphisms of the root system, and by the construction of the previous paragraph, gives rise to inner automorphisms of $\mathfrak{g}$. Outer automorphisms will arise from root system isomorphisms that are not in the Weyl group.

\paragraph{The $\mathfrak{sl}(N,\mathbb{C})$ case}
In the case of $\mathfrak{g} = \mathfrak{sl}(N,\mathbb{C})$ with $N \geq 3$ the root system isomorphisms that are not in the Weyl group are of the form $-w$ for $w \in W$. If we choose $c_{\alpha} = 1$ for all the simple roots, then on can generate automorphisms in all possible classes from $W \cup (-W)$. It should be noted that the class of an involutive Lie algebra automorphism associated to a given root system automorphism depends on the choice of the $c_{\alpha}$, as illustrated by the example below.

\paragraph{The $\mathfrak{sl}(4,\mathbb{C})$ example}
Let us illustrate this with the concrete example of $\mathfrak{g} = \mathfrak{sl}(4,\mathbb{C})$. We express root system automorphisms as matrices in the basis of the simple roots. Thus the corresponding Weyl group is the order $4!$ group generated by the three simple reflections
\begin{equation}
    \left(
\begin{array}{ccc}
 -1 & 1 & 0 \\
 0 & 1 & 0 \\
 0 & 0 & 1 \\
\end{array}
\right),\left(
\begin{array}{ccc}
 1 & 0 & 0 \\
 1 & -1 & 1 \\
 0 & 0 & 1 \\
\end{array}
\right),\left(
\begin{array}{ccc}
 1 & 0 & 0 \\
 0 & 1 & 0 \\
 0 & 1 & -1 \\
\end{array}
\right) \,.
\end{equation}
Out of these automorphisms, we now generate Lie algebra involutions. We first choose $c_\alpha = 1$. Then we find that exactly ten give rise to (inner) involutions of $\mathfrak{su}(4)$, and their type from Table \ref{tableRealForms} can be read from the multiplicity of the eigenvalue 1, which can be 15 (for $p=0$), 9 (for $p=1$) or 7 (for $p=2$). The identity corresponds to $p=0$ , six automorphisms correspond to $p=1$, namely
\begin{equation}
    \left(
\begin{array}{ccc}
 -1 & 1 & 0 \\
 0 & 1 & 0 \\
 0 & 0 & 1 \\
\end{array}
\right),\left(
\begin{array}{ccc}
 0 & -1 & 1 \\
 -1 & 0 & 1 \\
 0 & 0 & 1 \\
\end{array}
\right),\left(
\begin{array}{ccc}
 0 & 0 & -1 \\
 -1 & 1 & -1 \\
 -1 & 0 & 0 \\
\end{array}
\right)\,,
\end{equation}
\begin{equation}
  \left(
\begin{array}{ccc}
 1 & 0 & 0 \\
 0 & 1 & 0 \\
 0 & 1 & -1 \\
\end{array}
\right),\left(
\begin{array}{ccc}
 1 & 0 & 0 \\
 1 & -1 & 1 \\
 0 & 0 & 1 \\
\end{array}
\right),\left(
\begin{array}{ccc}
 1 & 0 & 0 \\
 1 & 0 & -1 \\
 1 & -1 & 0 \\
\end{array}
\right) \,,
\end{equation}
and three automorphisms correspond to $p=2$, namely
\begin{equation}
    \left(
\begin{array}{ccc}
 -1 & 1 & 0 \\
 0 & 1 & 0 \\
 0 & 1 & -1 \\
\end{array}
\right),\left(
\begin{array}{ccc}
 0 & -1 & 1 \\
 0 & -1 & 0 \\
 1 & -1 & 0 \\
\end{array}
\right),\left(
\begin{array}{ccc}
 0 & 0 & -1 \\
 0 & -1 & 0 \\
 -1 & 0 & 0 \\
\end{array}
\right) \,.
\end{equation}
Now we turn to the outer automorphisms, generated by $-W$. Here there are exactly six involutive outer automorphisms, and their type can be read from the multiplicity of the eigenvalue 1, which is 6 for type I and 10 for type II. We find four involutions of type I, namely
\begin{equation}
  \left(
\begin{array}{ccc}
 -1 & 0 & 0 \\
 0 & -1 & 0 \\
 0 & 0 & -1 \\
\end{array}
\right),  \left(
\begin{array}{ccc}
 0 & 1 & -1 \\
 1 & 0 & -1 \\
 0 & 0 & -1 \\
\end{array}
\right),\left(
\begin{array}{ccc}
 0 & 1 & -1 \\
 0 & 1 & 0 \\
 -1 & 1 & 0 \\
\end{array}
\right),\left(
\begin{array}{ccc}
 -1 & 0 & 0 \\
 -1 & 0 & 1 \\
 -1 & 1 & 0 \\
\end{array}
\right)
\end{equation}
and two of type II, namely
\begin{equation}
  \left(
\begin{array}{ccc}
 0 & 0 & 1 \\
 0 & 1 & 0 \\
 1 & 0 & 0 \\
\end{array}
\right) ,   \left(
\begin{array}{ccc}
 1 & -1 & 0 \\
 0 & -1 & 0 \\
 0 & -1 & 1 \\
\end{array}
\right) \,.
\end{equation}

Now we can do the same exercise with $c_\alpha = -1$. In that case, there are only 6 inner involutions generated by the Weyl group, two of them of type $p=1$ and four of them of type $p=2$. There are 10 outer involutions generated by $-W$, all of them of type $I$.

\paragraph{The flip involution} Let us focus on a particular element of $-W$, namely the \emph{flip} defined by
\begin{equation}
\label{definitionFlip}
    \alpha_i \rightarrow \alpha_{N-i}\,.
\end{equation}
When $N$ is odd, the flip is of course always of type I. On the other hand, it turns out that when $N$ is even, the flip generates an outer involutive automorphism of type I when we choose $c_\alpha = -1$, while it generated an outer involutive automorphism of type II when we choose $c_\alpha = +1$. This observation gives us a definition of the two groups $\widetilde{\mathrm{SU}}(N)_{\mathrm{I,II}}$ that just differs by a sign, namely, we use (\ref{definitionSUtilde}) where $\Theta_{\mathrm{I,II}}$ is the flip defined using for all simple root $\alpha$
\begin{equation}
\label{definitionc}
    c_\alpha \equiv c = \begin{cases} -1 & \textrm{ for type } \mathrm{I} \\
    +1 & \textrm{ for type } \mathrm{II} \\
    \end{cases}
\end{equation}
This is summarized in Table \ref{tableFlip}. It is easy to prove by recursion on the height\footnote{We recall that the height of a root $\alpha$, denoted by $\mathrm{ht}(\alpha)$, is the sum of its coefficients when expressed in the basis of simple roots. } of the root $\alpha$ that the extensions of the flip to the Lie algebra are defined by
\begin{equation}
\label{thetaIandII}
    \theta_{\mathrm{I,II}} \left(X_{\alpha} \right) = - (-c)^{\mathrm{ht}(\alpha)}X_{\phi (\alpha)} \, ,
\end{equation}
for any root $\alpha$ (in the case of simple roots, this reduces to (\ref{defcalpha})). The corresponding Lie group morphisms are called $\Theta_{\mathrm{I,II}}$.

\begin{table}
\begin{equation*}
    \begin{array}{|c|c|c|} \hline
        \textrm{Value of } c_\alpha & N \textrm{ odd} & N \textrm{ even} \\ \hline
         c_\alpha = +1  & \mathrm{I} & \mathrm{II} \\
         c_\alpha = -1  & -  & \mathrm{I} \\ \hline
    \end{array}
\end{equation*}
    \caption{Type of outer involutive automorphisms generated by the flip of the Dynkin diagram of $A_{N-1}$ as a function of the parity of $N$ and of the choice of the constant $c_{\alpha}$, taken to be the same for all the simple roots.}
    \label{tableFlip}
\end{table}

The fundamental representation is given by (\ref{defGtilde}), where the matrix $A$ is
\begin{equation}
\label{matrixA}
    A = \left(
    \begin{array}{cccccc}
         & & & & & 1 \\
         & & & & -c &  \\
         & & & (-c)^2 & &  \\
         & & \dots & & &   \\
         & \dots& & & &   \\
         (-c)^{N-1} & & & & &   \\
    \end{array}
    \right) \, .
\end{equation}
One checks that these matrices satisfy the symmetry properties encountered in section \ref{matrixrealization}.

\section{Representations, invariants and integration measure}\label{reps}

Having established the existence of the two groups $\widetilde{\mathrm{SU}}(N)_{\mathrm{I,II}}$, our next task is to study them. An aspect of primary interest are their representations, in particular the fundamental and the adjoint, as they will provide the basic building blocks to construct gauge theories based on $\widetilde{\mathrm{SU}}(N)_{\mathrm{I,II}}$. In turn, the possible invariants which can be constructed out of them will also play a role, as they will either enter the construction of the Lagrangian of the theories or because they will be identified with gauge-invariant operators. The latter can be systematically constructed by computing index-like (we will be more precise below) generating functions such as the Higgs branch Hilbert series or the Coulomb branch limit of the index, for which a necessary tool is the integration measure on these groups.

\subsection{Representations}

Let us begin with some general remarks, aiming at understanding the representations of $\widetilde{\mathrm{SU}}(N)_{\mathrm{I,II}}$ from those of $\mathrm{SU}(N)$. To that matter, we adapt the discussion of \cite{brocker2013representations} (section VI.7), regarding \emph{induced representations}. Let us define $\widetilde{G} = \widetilde{\mathrm{SU}}(N)$ and $G = \mathrm{SU}(N)$. We have $\widetilde{G}/G \simeq \mathbb{Z}_2$. For $k \in \mathbb{Z}_2$, we define $\Omega(k)$ the representation of $\widetilde{G}/G$,
\begin{eqnarray}
\Omega (k) &:& \widetilde{G}/G \times \mathbb{C} \rightarrow  \mathbb{C}  \\
&& (x,z) \mapsto (-1)^k z \nonumber\, .
\end{eqnarray}
Using the canonical projection $\widetilde{G} \rightarrow \widetilde{G}/G$, $\Omega (k)$ can also be seen as a representation of $\widetilde{G}$.

Now we consider two constructions:
\begin{itemize}
    \item From a representation $V$ of $\widetilde{G}$, one can construct other representations $V \otimes \Omega (k)$ of $\widetilde{G}$ for $k \in \widetilde{G}/G$. Note that $V\otimes\Omega(0)\cong V$.
    \item From a representation $U$ of $G$, one can construct other representations $U_x$ of $G$ for $x \in \widetilde{G}/G$. These are defined by the action of $G$ on $U_x$ given by $g \cdot u = \widetilde{g}\,g\,\widetilde{g}^{-1}u$ where $\widetilde{g}\in \widetilde{G}$ is such that $x \in g\,G$.
\end{itemize}
These representations can be partitioned into two types, according to Table \ref{tablerepresentations}. The reason for this classification is that induction and restriction relate representations of the same type. Now consider a representation of $\mathrm{SU}(N)$, which we call $U_0$, with Dynkin labels $[\lambda_1 , \cdots , \lambda_{N-1}]$, (this means that these are the coefficients of the highest weight in the basis of fundamental weights). Since we are working with the flip involution \eqref{definitionFlip}, the twisted representation by the non-trivial element of $\widetilde{G}/G$, $U_1$, has Dynkin labels $[\lambda_{N-1} , \cdots , \lambda_1]$. Therefore we are in type A if and only if $\lambda_i = \lambda_{N-i}$ for all $i$. As a consequence:
\begin{itemize}
    \item If an $\mathrm{SU}(N)$ representation $U$ has $\lambda_i = \lambda_{N-i}$ for all $i$, then the induced representation on $\widetilde{\mathrm{SU}}(N)$ is reducible and can be written $(V \otimes \Omega(0)) \oplus (V \otimes \Omega(1))$.
    \item If an $\mathrm{SU}(N)$ representation $U$ has $\lambda_i \neq \lambda_{N-i}$ for some $i$, then the induced representation on $\widetilde{\mathrm{SU}}(N)$ is irreducible (and is the same as the induced representation from $[\lambda_{N-1} , \cdots , \lambda_1]$).
\end{itemize}

\begin{table}
    \centering
    \begin{tabular}{|c|c|c|} \hline
         & Type A & Type B \\  \hline
    $G$-representations $U$ & all $U_x$ isomorphic & all $U_x$ distinct \\
    $\widetilde{G}$-representations $V$ & all $V \otimes \Omega (k)$ distinct & all $V \otimes \Omega (k)$ isomorphic \\ \hline
    \end{tabular}
    \caption{Types of representations related by induction and restriction. See theorem VI.7.3 of \cite{brocker2013representations}: If $U$ is a representation of $G$ of type A, the induced representation of $\widetilde{G}$ is $\textrm{ind}^{\widetilde{G}}\,U=\bigoplus_k V\otimes\Omega(k)$. If $U_x$, $x\in \widetilde{G}/G$ are of type B, they all induce the same representation on $\widetilde{G}$, $\text{ind}^{\widetilde{G}}\,U_x=V$.}
    \label{tablerepresentations}
\end{table}

For instance, we have
\begin{itemize}
    \item The fundamental $[1,0,\cdots,0]$ of $\mathrm{SU}(N)$ induces a unique irreducible representation of $\widetilde{\mathrm{SU}}(N)$. It has dimension $2N$.
    \item The adjoint $[1,0,\cdots ,0,1]$ of $\mathrm{SU}(N)$ induces a reducible representation of $\widetilde{\mathrm{SU}}(N)$, which decomposes into two irreducibles.
\end{itemize}

Let us now explicitly construct the fundamental and the adjoint representations, which will be relevant for our later purposes.

\subsubsection{The fundamental representation}\label{fund}

A particularly important representation will be the fundamental representation. It corresponds to the the matrix representation introduced in section \ref{matrixrealization}, which acts on a $2N$ dimensional complex space $\mathbb{C}^N\times \mathbb{C}^N$. Note that we may alternatively think of this space as $\mathbb{C}^N\times (\mathbb{C}^{\star})^N$, thus making explicit that $\widetilde{\mathrm{SU}}(N)_{\mathrm{I,II}}$ representations comprise a fundamental and antifundamental of the connected component $\mathrm{SU}(N)$. The elements of this space are of the form
\begin{equation}
\mathbf{Q}=\left(\begin{array}{c} \vec{x}\\ \vec{y} \end{array}\right)\,,\qquad \vec{x}=\left(\begin{array}{c}x_1\\ x_2\\ \vdots\\ x_N\end{array}\right)\,,\qquad \vec{y} =\left(\begin{array}{c}y_1\\ y_2\\ \vdots\\ y_N\end{array}\right) \, .
\end{equation}
It is useful to introduce a ``conjugate"
\begin{equation}
\overline{\mathbf{Q}}=\mathbf{Q}^T\,\mathbf{\Gamma_0}\,, \qquad \textrm{with} \qquad
\mathbf{\Gamma_0}=
\left(
\begin{array}{cc}
0 & \unity \\
-c \, \unity & 0
\end{array}
\right) \, .
\end{equation}
Then, for a generic $\widetilde{\mathbf{U}}\in \widetilde{G}$, $\mathbf{Q}$ and $\overline{\mathbf{Q}}$ transform as
\begin{equation}
\label{gaugetransform}
\mathbf{Q}\rightarrow \widetilde{\mathbf{U}} \,\mathbf{Q}\,,\qquad \overline{\mathbf{Q}}\rightarrow \overline{\mathbf{Q}}\,\widetilde{\mathbf{U}}^{\dagger}\,.
\end{equation}

\subsubsection{The adjoint representation}\label{adj}

Another very important representation for our purposes is the adjoint representation. Given the matrix representation in section \ref{matrixrealization}, an element $\mathbf{\Phi}$ in the adjoint representation is the $2N \times 2N$ block-diagonal matrix (recall that $\Phi \in \mathfrak{su}(N)$, so $\Phi^{\dagger} = \Phi$ and the Lie algebra automorphism --complex conjugation for hermitean generators-- is $\Phi \mapsto - \Phi^{\star}$)
\begin{equation}
    \mathbf{\Phi} = \left(\begin{array}{cc} \Phi & 0 \\ 0 & -\Phi^{\star} \end{array}\right)\,.
\end{equation}
Under $\widetilde{G}$ it transforms as
\begin{equation}
\label{adjtransf}
  \mathbf{\Phi}\rightarrow \widetilde{\mathbf{U}} \, \mathbf{\Phi} \,\widetilde{\mathbf{U}}^{\dagger}\,.
\end{equation}
For future purposes, it is interesting to note that
\begin{equation}
\label{Gamma0Phi}
    \mathbf{\Gamma_0} \, \mathbf{\Phi}^T\, \mathbf{\Gamma_0} = c \, \mathbf{\Phi} \, .
\end{equation}

Note that since one block is complex-conjugated of the other, the number of degrees of freedom is really $N^2-1$ as it should be for the adjoint. On the other hand, expressing the adjoint in this way turns out to be most convenient for our latter purposes of constructing gauge theories based on $\widetilde{\mathrm{SU}}(N)_{\mathrm{I,II}}$ due to the transformation properties expressed as \eqref{adjtransf}.

\subsection{Invariants}

Having explicitly constructed the fundamental and the adjoint representations, we now study the invariants which can be constructed out of them. To that matter, let us consider $F$ copies of the fundamental representation in addition to an adjoint representation. To set notation, we will denote $\widetilde{G}$ indices by $\widetilde{\alpha}$ with $\widetilde{\alpha}=1,\,\cdots,\, 2N$; and ``global symmetry indices" by $I$ with $I=1,\,\cdots,\,F$. To be explicit with the notation, the fundamentals will be $(\mathbf{Q}_I)^{\widetilde{\alpha}}$. Note that it follows that the indices of the conjugate are $(\overline{\mathbf{Q}}_I)_{\widetilde{\alpha}}$.

With the transformation rules described above for these representations, we may construct all possible group invariants made out of them. Let us stress that the list of such group invariants is infinite and we will not attempt for an exhaustive classification. Instead, we will focus on the ones which will be of uttermost relevance for our purposes. Indeed, we use a gauge-theoretic inspired naming with an eye on applications to gauge theories. Such most relevant invariants are
\begin{enumerate}
    \item Meson-like invariants: consider
    \begin{equation}
    \label{mesons}
        \mathbf{M}_{IJ}=(\overline{\mathbf{Q}}_I)_{\widetilde{\alpha}}\,(\mathbf{Q}_J)^{\widetilde{\alpha}}\equiv\overline{\mathbf{Q}}_I\,\mathbf{Q}_J\,.
        \end{equation}
        It is clear that such quantity is an invariant of the group action, using (\ref{gaugetransform}). Moreover, a short computation\footnote{$\mathbf{M}_{IJ} =\overline{\mathbf{Q}}_I\,\mathbf{Q}_J = ( \overline{\mathbf{Q}}_I\,\mathbf{Q}_J)^T = \mathbf{Q}_J^T \mathbf{\Gamma_0}^T \mathbf{Q}_I =-c \mathbf{Q}_J^T \mathbf{\Gamma_0} \mathbf{Q}_I  = -c  \, \mathbf{M}_{JI} \, .$ } shows that, as a $F\times F$ matrix
      \begin{equation}
    \label{symmetryMesons}
       \mathbf{M}_{IJ} = -c  \, \mathbf{M}_{JI} \, .
    \end{equation}
    \item Baryon-like invariants: introduce the $\epsilon$-like tensor $\Upsilon_{\widetilde{\alpha}_1\cdots\widetilde{\alpha}_N}$ such that
    \begin{equation}
        \Upsilon_{\widetilde{\alpha}_1\cdots\widetilde{\alpha}_N}=\begin{cases} \epsilon_{\widetilde{\alpha}_1\cdots\widetilde{\alpha}_N} \qquad {\rm if}\qquad \widetilde{\alpha}_i\in 1,\cdots N\,\,\,\forall i \\ \epsilon_{\widetilde{\alpha}_1\cdots\widetilde{\alpha}_N} \qquad {\rm if}\qquad \widetilde{\alpha}_i\in N+1,\cdots 2N\,\,\,\forall i \\ 0 \qquad{\rm otherwise}
        \end{cases} \qquad ;
    \end{equation}
    where $\epsilon_{\widetilde{\alpha}_1\cdots\widetilde{\alpha}_N}$ is the standard $\epsilon$-tensor in $\mathrm{SU}(N)$. Then we have the baryon-like invariants $\mathbf{B}_{I_1\cdots I_F}$ given by
    \begin{equation}
    \label{barions}
        \mathbf{B}_{I_1\cdots I_F}=(\mathbf{Q}_{I_1})^{\widetilde{\alpha}_1}\cdots (\mathbf{Q}_{I_F})^{\widetilde{\alpha}_F}\,\Upsilon_{\widetilde{\alpha}_1\cdots\widetilde{\alpha}_N}\,.
    \end{equation}
    Note that $\mathbf{B}_{I_1\cdots I_F}$ is completely antisymmetric on its $F$ indices.
    \item Superpotential-like invariants: consider
    \begin{equation}
    \label{Ws}
        \overline{\mathbf{Q}}_I\,\mathbf{\Phi}\,\mathbf{Q}_J\,.
        \end{equation}
       It is clear that such quantity is an invariant. Note that we may replace $\mathbf{\Phi}$ by $\mathbf{\Phi}^n$, since the $n$-power of an adjoint still transforms in the same way. Moreover, as a $F\times F$ matrix, we have
    \begin{equation}
    \label{SymmetryAdjMes}
         \overline{\mathbf{Q}}_I\, \mathbf{\Phi}\, \mathbf{Q}_J = c \, \overline{\mathbf{Q}}_J\, \mathbf{\Phi}\, \mathbf{Q}_I \, .
    \end{equation}
    \item Coulomb branch-like invariants: consider
    \begin{equation}
    \label{Coulomb}
        {\rm Tr}\,\mathbf{\Phi}^{2n}\,.
        \end{equation}
        It is clear that these quantities are invariant under the group transformations above.
\end{enumerate}

Note that these are ``holomorphic" invariants in that they do not make use of complex conjugation. On top of them, and explicitly using complex conjugation, we can construct the ``non-holomorphic" quantity (which we will dub K\"{a}hler-like)
    \begin{equation}
    \label{chiralAntichiral}
        \mathbf{Q}_I^{\dagger}\,\mathbf{Q}_J\,,
    \end{equation}
which is also invariant under the above transformations.

\subsection{The integration measures}

In this section, we consider only the case $N$ even (for $N$ odd, we refer to \cite{Bourget:2019phe}). In order to be able to compute index-like quantities for gauge theories based on $\widetilde{\mathrm{SU}}(N)_{\mathrm{I,II}}$, we need the integration measures on said groups. Recall that the standard way of defining the Haar measure of a \emph{connected} Lie group, grounded on the fact that conjugation of elements of the maximal torus of the group is surjective onto the full group, doesn't apply to our situation. Instead, to be able to integrate over the disconnected component of $\widetilde{\mathrm{SU}}(N)$ we use Lemma 2.1 of \cite{wendt2001weyl}, namely the fact that the map
\begin{align}
    \varphi:&\, \mathrm{SU}(N)/S_0(\Theta)\times S_0(\Theta)\to \mathrm{SU}(N)\Theta\\
    & (y S_0(\Theta) ,z)\quad\mapsto\quad yz\Theta y^{-1}\nonumber
\end{align}
where $S_0(\Theta)$ is the subgroup of the maximal torus of $\mathrm{SU}(N)$ left invariant by the involution $\Theta$, is surjective onto the component of $\widetilde{\mathrm{SU}}(N)$ disconnected from the identity. Therefore, we can use $\varphi$ as a change of variables, and turn the integration over $\mathrm{SU}(N)\Theta$ into one over $S_0(\Theta)$. The measure arises from the Jacobian of the change of variables,
\begin{align}
\label{determinant}
   \left. \det\left(\mathrm{d} \varphi\right)(y,z)=\det\left(  \text{Ad}( z \Theta)^{-1}-\text{Id}\right)\right|_{\mathfrak{su}(N)/\mathfrak{s_0(\theta)}} \, \ ,
\end{align}
where $\mathfrak{s_0(\theta)}$ is the Lie algebra of $S_0(\Theta)$. The Jacobian (\ref{determinant}) can be easily calculated from the data in the root system, since the involution $\Theta$ is completely defined by the flip $\phi$ (\ref{definitionFlip}) of the roots and the sign $c$ introduced in (\ref{definitionc}). As in \cite{Bourget:2019phe}, we use an adapted parametrization for the fugacities,
\begin{equation}
\label{defz}
   z^{\lambda} = 
       \left( \prod\limits_{i=1}^{\frac{N}{2}-1} z_i^{\lambda_i + \lambda_{N-i}}  \right) \left( \prod\limits_{i=1}^{\frac{N}{2}-1} z_{\frac{N}{2}+i}^{\lambda_i - \lambda_{N-i}} \right) z_{\frac{N}{2}}^{\lambda_{\frac{N}{2}}} 
\end{equation}
If a root $\alpha$ is fixed by $\phi$, the corresponding element of the Lie algebra $X_{\alpha}$ is transformed to $-(-c)^{\mathrm{ht}(\alpha)} X_{\alpha} = c X_{\alpha}$ since the height is necessarily odd, and it will contribute $(1-cz^{-\alpha})$ to the determinant (\ref{determinant}). On the other hand, if $\alpha$ is exchanged with $\phi(\alpha)$, their contribution will come from the determinant of the block matrix
\begin{equation}
\det\left(\begin{array}{cc}
        -1 & -(-c)^{\mathrm{ht}(\alpha)} z^{-\alpha}\\
        -(-c)^{\mathrm{ht}(\alpha)} z^{-\phi(\alpha)} & -1
    \end{array}\right) =  1-z^{-\alpha-\phi(\alpha)} \, ,
\end{equation}
where we have used (\ref{thetaIandII}).
In total, the integration measure is
\begin{equation}
\label{measure}
   \mathrm{d} \mu_{\mathrm{I,II}}^{-}(z) =   \prod\limits_{\alpha = \phi (\alpha)} \left( 1- c z^{-\alpha} \right)  \prod\limits_{\alpha \neq \phi (\alpha)} \left( 1-  z^{-(\alpha + \phi (\alpha))} \right)^{1/2} \prod\limits_{j=1}^{N/2} \frac{\mathrm{d} z_j}{2 \pi i z_j} \, .
\end{equation}
In (\ref{measure}), the products run over the positive roots. In the second product, the power $\frac{1}{2}$ takes care of the fact that each pair of roots is counted twice. The integration over $\widetilde{\mathrm{SU}}(N)_{\mathrm{I,II}}$ for $N$ even is then obtained by taking an average,
\begin{equation}
     \int_{\widetilde{\mathrm{SU}}(N)_{\mathrm{I,II}}}d\mu(X) f(X) = \frac{1}{2}\left( \int_{\mathrm{SU}(N)}d\mu^+(z)f(z) + \int_{\mathrm{SU}(N)\Theta_{\mathrm{I,II}}} d\mu^-_{\mathrm{I,II}}(z)f(\Theta_{\mathrm{I,II}}(z))  \right)\, ,
\end{equation}
where $f$ is a function defined on $\widetilde{\mathrm{SU}}(N)_{\mathrm{I,II}}$ which is invariant under conjugation and $d\mu^+$ is the standard Haar measure of $\mathrm{SU}(N)$.

\subsection{Real and pseudo-real representations}
\label{secFS}

Having constructd a group measure allows us to construct an indicator --the so-called Frobenius-Schur indicator-- sensible to the reality properties of the representations. This is very useful since, in Physics language, allows us to discern whether we have an orthogonal, unitary or symplectic global symmetry. We first quote the Frobenius-Schur theorem (see \cite{bump2004lie}, Theorem 43.1). Consider an irreducible representation $\rho$ of a compact group $G$, and compute the quantity
\begin{equation}
    \mathrm{FS}(\rho) = \int_G \chi_{\rho} (g^2) \mathrm{d} g \, ,
\end{equation}
where $\chi_{\rho}$ is the character of the representation. Then
\begin{eqnarray}
    \mathrm{FS}(\rho) = 1 &\Longleftrightarrow& \rho \textrm{ is real} \\
        \mathrm{FS}(\rho) = 0 &\Longleftrightarrow& \rho \textrm{ is complex} \\
            \mathrm{FS}(\rho) = -1 &\Longleftrightarrow& \rho \textrm{ is pseudo-real}
\end{eqnarray}

Using this, combined with the integration formula, we can investigate the properties of our representations. Let us focus on the fundamental representation of $\widetilde{\mathrm{SU}}(N)_{\mathrm{I,II}}$, for $N$ even. Using the measure (\ref{measure}), we evaluate
\begin{equation}
   \mathrm{FS}(\mathrm{Fund}) =  -c \,.
\end{equation}
This means that the fundamental representation is real in type I and pseudo-real in type II. This will have consequences in the next section, when we will study 4d $\mathcal{N}=2$ gauge theories with fundamental matter in hypermultiplets: the unitary global symmetry that exchanges copies of the fundamental is enhanced to
\begin{itemize}
    \item Symplectic global symmetry when the representations are real, i.e. in type I;
    \item Orthogonal global symmetry when the representations are pseudo-real, i.e. in type II.
\end{itemize}

\section{\texorpdfstring{Construction of $\mathcal{N}=2$ gauge theories}{Construction of N=2 gauge theories}}\label{gauge}

We now explicitly construct gauge theories based on $\widetilde{\mathrm{SU}}(N)_{\mathrm{I,II}}$ groups. For definiteness, we will construct 4d $\mathcal{N}=2$ SQCD-like theories.

A point which is worth emphasizing is that the $\widetilde{\mathrm{SU}}(N)_{\mathrm{I,II}}$ groups are not just the direct product of $SU(N)$ and the charge conjugation $\mathbb{Z}_2$. This bars the simple construction of general gauge theories based on $\widetilde{\mathrm{SU}}(N)_{\mathrm{I,II}}$ as the extra gauging of $\mathbb{Z}_2$ in a standard $SU(N)$ $\mathcal{N}=2$ theory, a procedure which would be tantamount to considering the direct product $SU(N)\times \mathbb{Z}_2$ which in general would not be consistent. An intuitive reason is that complex conjugation cannot be disjoint from gauge transformations since these are in general complex. The advantage of constructing the $\widetilde{\mathrm{SU}}(N)_{\mathrm{I,II}}$ groups is that this problem is \textit{ab initio} circumvented and hence the standard technology to construct gauge theories can be directly imported.

\subsection{Matter content}

The relevant multiplets to construct our theories are

\begin{itemize}
\item Vector multiplet: \newline
The vector multiplet contains, in 4d $\mathcal{N}=1$ language, a vector multiplet and a chiral multiplet in the adjoint. The latter will be described by an adjoint superfield which we will denote by $\mathbf{\Phi}$ with the transformation properties described in section \ref{adj}.
\item (Fundamental) Hypermultiplet: \newline
In order to construct hypermultiplets, let us take two chiral superfields, say $\mathbf{A}$ and $\mathbf{B}$, transforming in the fundamental representation as described in section \ref{fund}. Out of, say, $\mathbf{B}$, we can construct the corresponding $\overline{\mathbf{B}}$. Let us now consider constructing a chiral superfield out of $\mathbf{A}$, and another chiral superfield out of $\mathbf{B}$, but instead taking its barred cousin, \textit{i.e.} $\overline{\mathbf{B}}$. Let us insist once again that \emph{both} $\mathbf{A}$ and $\overline{\mathbf{B}}$ are chiral superfields of the \emph{same} chirality. Thus, we may construct a hypermultiplet out of them, \textit{i.e.}
\begin{equation}
\mathcal{H}=(\mathbf{A},\,\overline{\mathbf{B}})\,.
\end{equation}
Note that both $\mathbf{A}$ and $\overline{\mathbf{B}}$ provide $2N$ degrees of freedom, so that $\mathcal{H}$ contains 4N degrees of freedom.

\end{itemize}

Having described the basic ingredients, the construction of the Lagrangian follows the standard techniques in supersymmetric gauge theories. The kinetic terms will come from a K\"{a}hler potential. A natural candidate is the ``non-holomorphic" invariant \eqref{chiralAntichiral} above, \textit{i.e.} (we quote the free case; we will comment on the gauged version below)
\begin{equation}
K=\mathbf{A}^{\dagger}\,\mathbf{A}+\overline{\mathbf{B}}\,\overline{\mathbf{B}}^{\dagger}\, .
\end{equation}

Let us now turn to the superpotential $W$. Since it is an integration over half of superspace, it can only involve the chiral fields in $\mathcal{H}$. Assuming a number $F$ of hypermultiplets, the natural $W$ can be constructed out of \eqref{SymmetryAdjMes},
\begin{equation}
\label{W}
W=\overline{\mathbf{B}}_J\,\mathbf{\Phi}\,\mathbf{A}_I\,\mathbf{G}^{IJ} \, .
\end{equation}
with $\mathbf{G}^{IJ}$ a suitable matrix of couplings which would be fixed by the requirement of $\mathcal{N}=2$ SUSY.

So far we have evaded the gauge sector. By construction, only the part of the gauge group connected to the identity will contribute with a field in the Lagrangian, while the disconnected part of the gauge group will appear as a superselection rule (see \cite{Krauss:1988zc,Preskill:1990bm} for early discussions, and \cite{Argyres:2016yzz} for a more recent account). Thus, the vector multiplet will be the standard one associated to the gauge transformations in the $\mathrm{SU}(N)$ part of $\widetilde{\mathrm{SU}}(N)$.

\subsection{Smaller representations}

So far we have assumed $\mathbf{A}\ne \mathbf{B}$. But nothing prevents us from taking $\mathbf{A}=\mathbf{B}=\mathbf{Q}$. In this case, the hypermultiplet becomes $\mathcal{H}=(\mathbf{Q},\,\overline{\mathbf{Q}})$. Note that this cannot be done with a standard $\mathrm{SU}(N)$ hypermultiplet: indeed, if we want to construct invariants of $\mathrm{SU}(N)$ we need to consider a hypermultiplet $(Q,\,\tilde{Q})$ with $Q$ a fundamental of $\mathrm{SU}(N)$ and $\tilde{Q}$ an antifundamental. If we wanted ``$Q=\tilde{Q}$", we would have to set $\tilde{Q}\sim Q^{\star}$, and hence it would be a chiral superfield of the other chirality. The crucial difference is now that in the fundamental of $\widetilde{\mathrm{SU}}(N)_{\mathrm{I,II}}$ there is both the $\mathbf{N}$ and the $\overline{\mathbf{N}}$ of the connected $\mathrm{SU}(N)$ part, and the construction of the second element of the hyper --the equivalent to $\tilde{Q}$-- does not involve complex conjugation but rather a simple transposition, and hence does not change the chirality of the superfield. Since the degrees of freedom are half of the standard hyper, it would be perhaps more appropriate to call this $2N$ dimensional representation a half-hypermultiplet (note that in fact this is the same number of dof. as a full hypermultiplet of $SU(N)$).

All in all, we can write the theory for $F$ half-hypermultiplets. The $W$ is just the obvious particularization of \eqref{W}, \textit{i.e.}
\begin{equation}
\label{Whalf}
W=\overline{\mathbf{Q}}_J\,\mathbf{\Phi}\,\mathbf{Q}_I\,\mathbf{G}^{IJ}
\end{equation}
and using the symmetry property \eqref{SymmetryAdjMes} fixes the matrix $\mathbf{G}$ to be either symmetric or antisymmetric.
This allows us to immediately read the global symmetry of the theory:
\begin{itemize}
\item $\widetilde{\mathrm{SU}}(N)_{\mathrm{I}}$: $\mathbf{G}$ is antisymmetric. The global symmetry is  $\mathrm{Sp}(\frac{F}{2})$.
\item $\widetilde{\mathrm{SU}}(N)_{\mathrm{II}}$: $\mathbf{G}$ is symmetric. The global symmetry is  $\mathrm{SO}(F)$.
\end{itemize}
This is in perfect agreement with the result derived using the Frobenius-Schur indicator in section \ref{secFS}, and it will be confirmed by the explicit computation of the Higgs branch Hilbert series. Moreover, it is also suggested by table \ref{tableSymSpaces} -- the $\widetilde{\mathrm{SU}}(N)_{\mathrm{I,II}}$ behaves in this respect as its subgroup $K$ would. If $K$ is of orthogonal type, then the global symmetry will be symplectic, and vice versa.

Note that for the type I extensions the case of odd $F$ is not well-defined. The issue is manifest in the simplest case of $F=1$, where it is simply impossible to write a non-vanishing $W$. Since for any odd $F$ one can write $F=2f+1$, this very same argument suggests that type I theories with odd number of flavors do not exist as a $\mathcal{N}=2$ theories. In the following we will restrict our attention to even $F$ for type I theories.

\subsection{Dynamics}

In the following we will be interested in SCDQ theories with $\widetilde{SU}(N)_{\mathrm{I,II}}$ gauge group and $F$ fundamental half-hypers. As discussed above, the vector multiplet only contains a gauge field for the connected part of the gauge symmetry, while the disconnected part only enters as a superselection rule. As a consequence, the Lagrangian of the theory is just identical to that of its $SU(N)$ SQCD cousin. Hence, the Feynman rules will just be the same, and consequently, all local Physics will be identical to that of SQCD with the only extra addition that one has to impose the constraints arising from gauge invariance under the disconnected part of the gauge group (see \textit{e.g.} \cite{Krauss:1988zc,Preskill:1990bm,Argyres:2016yzz}).

An important consequence of these observations is that all (local) anomalies are just identical to those in SQCD, which, in particular, implies that pure gauge anomalies automatically vanish (a consequence of being a non-chiral theory). Note however that in general there may be 't Hooft anomalies associated to global symmetries (including mixed gauge-$U(1)_R$ anomalies, which vanish in the conformal case). In addition, there may be anomalies associated to the disconnected part of the gauge group. It would be very interesting to undertake a detailed analysis of this point.

Another very important consequence is that the $\beta$ function will just be the same as in SQCD. Thus, in particular we can tune $N$ and $F$ and restrict to well-behaved 4d QFT's. In particular, we can choose $N$ and $F$ so that our theories become conformal. This will be the most interesting case, since the gauge dynamics will greatly simplify due to the absence of a strong coupling scale and the full power of conformal invariance will provide very useful tools to analyze the theories. In particular, by means of the SCI we can study their spectrum in both the Coulomb and Higgs branches as we will do below. Note however that the Higgs branch is non-renormalized \cite{Argyres:1996eh}, and thus when, studying the Higgs branch, the requirements on $N$ and $F$ may be dropped (more on this below).

\section{The spectrum of the theory}\label{spectrum}

One aspect of basic interest is the operator content of the theories based on $\widetilde{G}$ and their relations. As discussed above, we can restrict to well-defined QFT's by choosing $N$ and $F$ so that the theory is at least asymptotically free. Nevertheless, in order to avoid the complicated gauge dynamics associated to the strong coupling scale of the gauge group, we can further focus on SCFT's. In that case, due to superconformal invariance, we have the powerful tool of the SCI to analyze the operator spectrum of the theory. While the full index is a complicated function, in particular limits it simplifies and allows to study in detail both the Coulomb branch and the Higgs branch of the theory.

Regarding the Coulomb branch, we can study the operator content through the so called Coulomb branch limit of the superconformal index \cite{Gadde:2011uv}.

In turn, for the Higgs branch, we can consider the Hall-Littlewood limit of the index. On general grounds, for a theory corresponding to a quiver with no loops, it is clear that the computation of such Hall-Littlewood limit of the index coincides with the computation of the Higgs branch Hilbert series, which is a counting of gauge-invariant operators made out of hypermultiplets \cite{Gaiotto:2012uq}.\footnote{The way this comes about is as follows: for a theory such as SQCD, the vector multiplet contribution to the index is through the gaugino, and it precisely coincides with the would-be contribution of the $F$-term constraint to the Hilbert series. On the other hand, the hypermultiplet contribution is just identical in both the Hall-Littlewood limit of the index and the Hilbert series.} Note however that, due to supersymmetry, the Higgs branch remains classical \cite{Argyres:1996eh}. Hence the computation of the Higgs branch Hilbert series using the classical Lagrangian even beyond the conformal window provides us a sensible description of the Higgs branch in the full quantum theory. Thus, when computing the Higgs branch Hilbert series, we will not restrict ourselves to theories in the conformal window. To be specific, we will consider below theories with gauge group $\widetilde{\mathrm{SU}}(N)_{\mathrm{I,II}}$ with $F$ half-hypermultiplets (in the sense discussed above) only when $F \geq 2(N-1)$. The $F<2N$ region is more difficult to study in part because at a generic point on the Higgs branch, the gauge group may not be completely Higgsed, which means that we can not use the letter-counting formula (\ref{eq:Hb}). A complete treatment of that question has appeared in \cite{future} for gauge groups $\mathrm{SU}(N)$, but the extension of the techniques used in \cite{future} to disconnected gauge groups remains an open problem.

\subsection{Warm-up: the free theory}

Let us first consider the free theory by sending the Yang-Mills coupling to zero. The spectrum of the theory will consist of gauge-invariant operators (as Gauss' law is kept as a constraint) with no other relation. Focusing on the Higgs branch --\textit{i.e.} on operators made out of hypermultiplet fields--, to lowest order the gauge invariants are the mesons $\mathbf{M}_{IJ}$ (we assume $N$ big enough so that baryon-like operators appear at high dimensions), which are either a symmetric (for $\widetilde{\mathrm{SU}}(N)_{\mathrm{I}}$) or an antisymmetric (for $\widetilde{\mathrm{SU}}(N)_{\mathrm{II}}$) $F\times F$ matrix. Thus, introducing a fugacity $t$ to count dimensions,  we should expect the first non-trivial contribution a (unrefined) partition function counting operators to be $t^2\,\frac{F\,(F\pm 1)}{2}$.\footnote{As argued above, the Hall-Littlewood limit index of the index coincides with the Higgs branch Hilbert series for the theories at hand. Nevertheless, strictly speaking, if though as the Hall-Littlewood limit of the index, the contribution of vector multiplet and hypermultiplet comes weigthed by a fugacity $\tau^R$. However, since the operators to count satisfy the BPS bound $\Delta=2R$, the difference between Higgs branch Hilbert series and Hall-Littlewood limit of the index is just a simple redefinition of the fugacity $t^2\leftrightarrow \tau$.}

Let us consider the next order $t^4$. For definitness, say we have odd $N$ --so we have $\widetilde{\mathrm{SU}}(N)_{\mathrm{I}}$. In that case, $\mathbf{M}$ is a symmetric matrix and hence has $d_S=\frac{F\,(F+1)}{2}$ entries. To order $t^4$ we will have the symmetrized product of those, \textit{i.e.} $\frac{d_S\,(d_S+1)}{2}$. In turn, while at order $t^2$, the $d_A=\frac{F\,(F-1)}{2}$ antisymmetric pieces of $\mathbf{M}$ are projected out, their symmetrized squares, \textit{i.e.} $\frac{d_A\,(d_A+1)}{2}$, survive at order $t^4$. Hence the $t^4$ coefficient is expected to be
\begin{equation}
\frac{1}{2}\Big[\frac{F\,(F+1)}{2}\,\Big(\frac{F\,(F+1)}{2}+1\Big)\Big]+\frac{1}{2}\Big[\frac{F\,(F-1)}{2}\,\Big(\frac{F\,(F-1)}{2}+1\Big)\Big]=\frac{F^2\,(F^2+3)}{4}\, .
\end{equation}
Note that, for $\widetilde{\mathrm{SU}}(N)_{\mathrm{II}}$ the roles of symmetric and antisymmetric are exchanged. Nevertheless this has no effect on the $t^4$ coefficient.  Hence, all in all, we expect
\begin{itemize}
\item $\widetilde{\mathrm{SU}}(N)_{\mathrm{I}}$:
\begin{equation}\label{eq:naiveoddN}
\textrm{HS}_{\mathrm{I}}^{\rm free}(t)=1+\frac{F\,(F+1)}{2}\,t^2+\frac{F^2\,(F^2+3)}{4}\,t^4+o(t^4)\, ;
\end{equation}
\item $\widetilde{\mathrm{SU}}(N)_{\mathrm{II}}$:
\begin{equation}
\textrm{HS}_{\mathrm{II}}^{\rm free}(t)=1+\frac{F\,(F-1)}{2}\,t^2+\frac{F^2\,(F^2+3)}{4}\,t^4+o(t^4)\, .
\end{equation}
\end{itemize}
Note that this implicitly assumes $N$ large enough. Indeed, if $N\leq 4$, the baryon would contribute to the order $t^4$. As we will see below, indeed the integration formula allows us to recover this expectation from the computation of the Higgs branch Hilbert series for the free theory.

Note that the $t^2$ term is somewhat special, in that its contributions come from scalars in conserved current multiplets (moment maps of the global symmetry). Since these are, by construction, in the adjoint representation of the global symmetry, the coefficient of $t^2$ provides a cross-check of the global symmetry of the theory. Indeed, for type $\mathrm{I}$ that coefficient coincides with the dimension of the adjoint of $\mathrm{Sp}(\frac{F}{2})$, while for type $\mathrm{II}$ it coincides with the dimension of the adjoint of $\mathrm{SO}(F)$.

Using the measure on the groups developed above we can cross-check (and extend to arbitrary order) the expectations above. The (free theory) Higgs branch Hilbert series reads
\begin{equation}
\label{eq:ring}
    \textrm{HS}^{\text{free}}_{(N,F)}(t)= \int_{G} d \eta_{G}(X) \frac{1}{\textrm{det}(1-t\Phi_{\mathrm{Fund}}(X))^{F}}\, .
\end{equation}
As discussed above, the integral splits into the sum of the connected and disconnected part, and the measures are the ones found in section \ref{reps}. It is then easy to show that indeed the expectation above for the first few terms is recovered. In order not to clutter the presentation, as an example, we quote the results for $\widetilde{\mathrm{SU}}(N)_{\mathrm{I}},$ and $F=2,4,6$ \footnote{The plethystic exponential (PE) of a function $f(x)$ such that $f(0)=0$ is defined as
\begin{equation*}
\textrm{PE}[f(x)]=\mathrm{exp} \left( \sum_{n=1}^{\infty}\frac{f(x^n)}{n} \right) \,.
\end{equation*}
}
\begin{align*}
    \textrm{HS}^{\text{free}}_{(4,2)}(t)=& \textrm{PE}[3t^2+t^4] = 1 + 3t^2 +7t^4 + o\left(t^4\right) \, \ , \\[4pt]
    \textrm{HS}^{\text{free}}_{(4,4)}(t)= & \frac{1-t^2+16t^4-10t^6+25t^8-5t^{10}+6t^{12}}{\left(1-t^2\right)^{17} \left(1+t^2\right)^6} = 1+10 t^2+77 t^4 +o\left(t^{4}\right)\, , \\[4pt]
    \textrm{HS}^{\text{free}}_{(4,6)}(t)=& \frac{1}{\left(1-t^2\right)^{33} \left(1+t^2\right)^{14}}\Big(1+2 t^2+124 t^4+435 t^6+3393 t^8+11034 t^{10}+38282 t^{12}+\\
    & 91513 t^{14}+195923 t^{16}+326359 t^{18}+476999 t^{20}+554635 t^{22}+569026 t^{24}+\\
    & 465194 t^{26}+334666 t^{28}+190410 t^{30}+95283 t^{32}+35694 t^{34}+12626 t^{36}+2599 t^{38}+\\
    & 734 t^{40}+45 t^{42}+15 t^{44}\Big)= 1+21 t^2+366 t^4 + o\left(t^{4}\right)\, \ ,
    \end{align*}
    and the results for $N=6$ and $F=2,4,6,8$:
    \begin{align*}
    \textrm{HS}^{\text{free}}_{(6,2)}(t)=&\textrm{PE}[3t^2+t^4]=1+3 t^2+7 t^4 +o\left(t^4\right)\, \ , \\
    \textrm{HS}^{\text{free}}_{(6,4)}(t)=&1+10 t^2+76 t^4 + o\left(t^4\right)\, \ ,\\
    \textrm{HS}^{\text{free}}_{(6,6)}(t)=&1+21 t^2+351 t^4 +o\left(t^4\right)\, \ , \\
    \textrm{HS}^{\text{free}}_{(6,8)}(t)=&1+36 t^2+1072 t^4 +o\left(t^4\right) \, \ .
\end{align*}
We observe that the first three terms in each of these examples match the expected result given by (\ref{eq:naiveoddN}), once the additional baryons that appear in the $t^4$ term for $N=4$, $F\ge 4$ are taken into account. In particular, this provides a confirmation of our expectations on the global symmetry to add to the computation of the Frobenius-Schur indicator as described above.

Furthermore, while both the numerator of the Hilbert series for the component connected with the identity  and the numerator of the Hilbert series for the component not connected with the identity are palindromic in the above examples, in general the full Hilbert series $\textrm{HS}_{(N,F)}^{\text{free}}(t)$ has not a palindromic numerator. Note however that in the free limit we are considering there is no \textit{a priori} reason for the Hilbert series to be palindromic (for instance, upon removing the $W$ the theory is effectively not even $\mathcal{N}=2$). Moreover, in general, the ring of invariants as a quite involved structure and the corresponding Highest Weights Generating function (HWG) \cite{Hanany:2014dia} does not seem to be given by a complete intersection.

\subsection{The full theory: Coulomb branch operators}

The Coulomb branch index is a counting of operators on the Coulomb branch of a CFT, and thus can be thought as a Hilbert series for the Coulomb branch. Note that the hypermultiplets only enter this computation through ensuring that we have a CFT, but otherwise they are blind to the computation of the Coulomb branch index. Thus, we will assume the matter content to be such that the theory has vanishing beta functions. From the transformation properties of the adjoint representations described above, it is clear that for either $\widetilde{\mathrm{SU}}(N)_{\mathrm{I,II}}$ the Coulomb branch will only count operators of the form ${\rm Tr}\,\Phi^{2n}$. Thus it is clear that \cite{Bourget:2019phe}
\begin{equation}
\label{CoulombHS}
    {\rm HS}^{C}_{N}(t)=\frac{1}{2}\Big[\prod_{n=2}^{N}\frac{1}{1-t^2}-\prod_{n=2}^{N}\frac{1}{1-(-t)^i}\Big]\,.
\end{equation}
This can be explicitly verified using the integration formula. On general grounds, for a theory with gauge group $G$, the Coulomb branch index (or Coulomb branch Hilbert series) reads
\begin{equation}
\label{eq:hsCB}
    \textrm{HS}^{C}_{N}(t) = \int_{G} d \eta_{G}(X) \frac{1}{\textrm{det}(1-t\Phi_{\textrm{Adj}}(X))}\, \ ,
\end{equation}
Using the formula (\ref{eq:hsCB}) we get
\begin{align}
&  \textrm{HS}^{C}_{4}(t) = \textrm{PE}[t^2+t^4+t^6]\, \ , \\
&  \textrm{HS}^{C}_{6}(t) = \textrm{PE}[t^2+t^4+2t^6+t^8+t^{10}-t^{16}]\, \ ,
\end{align}
which indeed agrees with \eqref{CoulombHS}.

Note that eq.\eqref{CoulombHS} shows that the Coulomb branch is identical as a complex variety for both $\widetilde{\mathrm{SU}}(N)_{\mathrm{I,II}}$. Moreover, it immediately follows that both families of theories provide explicit examples of consistent $\mathcal{N}=2$ QFT's with non-freely generated Coulomb branches, thus extending \cite{Bourget:2019phe}.

\subsection{The full theory: Higgs branch operators}

Let us now look to the operators in the Higgs branch for the $\mathcal{N}=2$ theories. Before delving in a full computation of the generating function of such operators, let us first obtain by hand the lowest lying such operators. To that matter we now need to additionally mod out by the F-terms. Note first that the F-terms can be computed by forgetting the vanishing trace requirement on the adjoint and adding a Lagrange multiplier $\lambda\,{\rm Tr}\,\mathbf{\Phi}$ to the $W$. Then, the F-terms are essentially
\begin{equation}
\label{F}
\mathbf{Q}_I\,\overline{\mathbf{Q}}_J\,\mathbf{G}^{IJ}=\mathbf{1}\,,
\end{equation}
where the free indices are in color space.

It is clear that the $F$ terms will enter first at order $t^4$. Thus, the coefficient of $t^2$ is just like in the free theory, and hence the same comment on the fact that it dictates the global symmetry of the theory applies. In turn, at order $t^4$ we need to take $F$ terms into account. Eq.\eqref{F}  essentially means that, when squaring $\mathbf{M}$ to construct the terms contributing to $t^4$, one combination of them, times a antisymmetric $F\times F$ matrix ($\widetilde{\mathrm{SU}}(N)_{\mathrm{II}}$) or symmetric matrix ($\widetilde{\mathrm{SU}}(N)_{\mathrm{I}}$) can be dropped. Hence, we should expect the $t^4$ term in the $\widetilde{\mathrm{SU}}(N)_{\mathrm{II}}$ case to be that of the free theory minus $\frac{F\,(F+1)}{2}$; while for $\widetilde{\mathrm{SU}}(N)_{\mathrm{I}}$ it should be that of the free theory minus $\frac{F\,(F-1)}{2}$. That is, we expect
\begin{itemize}
\item $\widetilde{\mathrm{SU}}(N)_{\mathrm{I}}$:
\begin{equation}
\textrm{HS}^{\mathrm{I}}(t)=1+\frac{F\,(F+1)}{2}\,t^2+\frac{F\,(F+1)}{2}\Big(\frac{F\,(F-1)}{2}+1\Big)\,t^4+\, o(t^4) \, \  ;
\end{equation}
 \item $\widetilde{\mathrm{SU}}(N)_{\mathrm{II}}$:
\begin{equation}
\textrm{HS}^{\mathrm{II}}(t)=1+\frac{F\,(F-1)}{2}\,t^2+\frac{F\,(F-1)}{2}\Big(\frac{F\,(F+1)}{2}+1\Big)\,t^4+\, o(t^4) \, \ .
\end{equation}
\end{itemize}

Just like in the free case, we can explicitly test this expectation and extend it to arbitrary orders in $t$ by explicitly computing the Higgs branch Hilbert series (recall, identical to the Hall-Littlewood limit of the index) using the Haar measure and the technology developed above. It generically reads
\begin{equation}
\label{eq:Hb}
 \textrm{HS}_{(N,F)}(t;q_i) =   \int_{G} d \eta_{G}(X) \frac{\textrm{det}(1-t^2\Phi_{\mathrm{Adj}}(X))}{\textrm{det}(1-t[1,0,...,0] \times \Phi_{\mathrm{Fund}}(X))} \, \ ,
\end{equation}
where the $\{q_i\}$ are set of global symmetry fugacity and $[1,0,...,0]$ is the Dynkin label for the fundamental representation of the global symmetry group.

In order to give a flavor of the computation, let us make explicit the ingredients in \eqref{eq:Hb} in the simplest example where the two outer involutions $\Theta_{\mathrm{I}}$ and $\Theta_{\mathrm{II}}$ are different, which is $\mathrm{SU}(4)$. Let's begin by choosing the following basis for the $\mathfrak{su}(4)$ Lie-algebra
\begin{eqnarray}
  &  \left\{ h_1, h_2, h_3,  X_{\alpha_1}, X_{\alpha_1+\alpha_2},  X_{\alpha_2}, X_{\alpha_1+\alpha_2+\alpha_3}, X_{\alpha_2+\alpha_3 } ,  X_{\alpha_3 } ,   \right. \\  & \left. X_{-\alpha_1}, X_{-\alpha_1-\alpha_2},  X_{-\alpha_2}, X_{-\alpha_1-\alpha_2-\alpha_3}, X_{-\alpha_2-\alpha_3 } ,  X_{-\alpha_3 }\right\} \nonumber
 \ ,
\end{eqnarray}
where the $h_i$, for $i=1,2,3$, denote generators of the Cartan subalgebra, while the $\alpha_i$ are the associated simple roots. For the type II extension, according to the discussion in section \ref{sectionliftLiealg}, the flip involution will act in the different representations of the Lie algebra as
\begin{equation}
    \Phi_{\mathrm{Adj}}(\Theta_{\mathrm{II}})=
\left(
\begin{array}{ccccccccccccccc}
 0 & 0 & 1 & 0 & 0 & 0 & 0 & 0 & 0 & 0 & 0 & 0 & 0 & 0 & 0 \\
 0 & 1 & 0 & 0 & 0 & 0 & 0 & 0 & 0 & 0 & 0 & 0 & 0 & 0 & 0 \\
 1 & 0 & 0 & 0 & 0 & 0 & 0 & 0 & 0 & 0 & 0 & 0 & 0 & 0 & 0 \\
 0 & 0 & 0 & 0 & 0 & 0 & 0 & 0 & 1 & 0 & 0 & 0 & 0 & 0 & 0 \\
 0 & 0 & 0 & 0 & 0 & 0 & 0 & -1 & 0 & 0 & 0 & 0 & 0 & 0 & 0 \\
 0 & 0 & 0 & 0 & 0 & 1 & 0 & 0 & 0 & 0 & 0 & 0 & 0 & 0 & 0 \\
 0 & 0 & 0 & 0 & 0 & 0 & 1 & 0 & 0 & 0 & 0 & 0 & 0 & 0 & 0 \\
 0 & 0 & 0 & 0 & -1 & 0 & 0 & 0 & 0 & 0 & 0 & 0 & 0 & 0 & 0 \\
 0 & 0 & 0 & 1 & 0 & 0 & 0 & 0 & 0 & 0 & 0 & 0 & 0 & 0 & 0 \\
 0 & 0 & 0 & 0 & 0 & 0 & 0 & 0 & 0 & 0 & 0 & 0 & 0 & 0 & 1 \\
 0 & 0 & 0 & 0 & 0 & 0 & 0 & 0 & 0 & 0 & 0 & 0 & 0 & -1 & 0 \\
 0 & 0 & 0 & 0 & 0 & 0 & 0 & 0 & 0 & 0 & 0 & 1 & 0 & 0 & 0 \\
 0 & 0 & 0 & 0 & 0 & 0 & 0 & 0 & 0 & 0 & 0 & 0 & 1 & 0 & 0 \\
 0 & 0 & 0 & 0 & 0 & 0 & 0 & 0 & 0 & 0 & -1 & 0 & 0 & 0 & 0 \\
 0 & 0 & 0 & 0 & 0 & 0 & 0 & 0 & 0 & 1 & 0 & 0 & 0 & 0 & 0 \\
\end{array}
\right) \, \ ,
\end{equation}
\begin{equation}
    \Phi_{\mathrm{Fund}}(\Theta_{\mathrm{II}})=
\left(
\begin{array}{cccccccc}
 0 & 0 & 0 & 0 & 0 & 0 & 0 & 1 \\
 0 & 0 & 0 & 0 & 0 & 0 & -1 & 0 \\
 0 & 0 & 0 & 0 & 0 & 1 & 0 & 0 \\
 0 & 0 & 0 & 0 & -1 & 0 & 0 & 0 \\
 0 & 0 & 0 & -1 & 0 & 0 & 0 & 0 \\
 0 & 0 & 1 & 0 & 0 & 0 & 0 & 0 \\
 0 & -1 & 0 & 0 & 0 & 0 & 0 & 0 \\
 1 & 0 & 0 & 0 & 0 & 0 & 0 & 0 \\
\end{array}
\right)\, \ .
\end{equation}

The measures for the connected and non-connected component with the identity have already been discussed in \cite{Bourget:2019phe}, and read
\begin{align}
    \mathrm{d} \mu_{4}^+(z)&=\frac{\mathrm{d} z_1}{2\pi i z_1}\frac{\mathrm{d} z_2}{2\pi i z_2}\frac{dz_3}{2\pi i z_3}\left(1-\frac{z_1^2z_3^2}{z_2}\right)\left(1-z_2z_3^2\right)\left(1-\frac{z_2^2}{z_1^2}\right)\left(1-z_1^2\right)\left(1-\frac{z_2}{z_3^2}\right)\left(1-\frac{z_1^2}{z_2z_3^2}\right) \, , \\
    \mathrm{d} \mu_{4,\mathrm{II}}^{-}(z)&=\frac{\mathrm{d} z_1}{2\pi i z_1}\frac{\mathrm{d} z_2}{2\pi i z_2}\left(1-\frac{z_1^4}{z_2^2}\right)\left(1-z_2^2\right)\left(1-\frac{z_2^2}{z_1^2}\right)\left(1-z_1^2\right) \, \ .
\end{align}

The second involutive outer automorphism is completely analogous except for the sign $c$ introduced in \eqref{definitionc}. The matrix $\Phi_{\mathrm{Adj}}(\Theta_{\mathrm{I}})$ reads
\begin{equation}
\Phi_{\mathrm{Adj}}(\Theta_I)=
\left(
\begin{array}{ccccccccccccccc}
 0 & 0 & 1 & 0 & 0 & 0 & 0 & 0 & 0 & 0 & 0 & 0 & 0
   & 0 & 0 \\
 0 & 1 & 0 & 0 & 0 & 0 & 0 & 0 & 0 & 0 & 0 & 0 & 0
   & 0 & 0 \\
 1 & 0 & 0 & 0 & 0 & 0 & 0 & 0 & 0 & 0 & 0 & 0 & 0
   & 0 & 0 \\
 0 & 0 & 0 & 0 & 0 & 0 & 0 & 0 & -1 & 0 & 0 & 0 & 0
   & 0 & 0 \\
 0 & 0 & 0 & 0 & 0 & 0 & 0 & -1 & 0 & 0 & 0 & 0 & 0
   & 0 & 0 \\
 0 & 0 & 0 & 0 & 0 & -1 & 0 & 0 & 0 & 0 & 0 & 0 & 0
   & 0 & 0 \\
 0 & 0 & 0 & 0 & 0 & 0 & -1 & 0 & 0 & 0 & 0 & 0 & 0
   & 0 & 0 \\
 0 & 0 & 0 & 0 & -1 & 0 & 0 & 0 & 0 & 0 & 0 & 0 & 0
   & 0 & 0 \\
 0 & 0 & 0 & -1 & 0 & 0 & 0 & 0 & 0 & 0 & 0 & 0 & 0
   & 0 & 0 \\
 0 & 0 & 0 & 0 & 0 & 0 & 0 & 0 & 0 & 0 & 0 & 0 & 0
   & 0 & -1 \\
 0 & 0 & 0 & 0 & 0 & 0 & 0 & 0 & 0 & 0 & 0 & 0 & 0
   & -1 & 0 \\
 0 & 0 & 0 & 0 & 0 & 0 & 0 & 0 & 0 & 0 & 0 & -1 & 0
   & 0 & 0 \\
 0 & 0 & 0 & 0 & 0 & 0 & 0 & 0 & 0 & 0 & 0 & 0 & -1
   & 0 & 0 \\
 0 & 0 & 0 & 0 & 0 & 0 & 0 & 0 & 0 & 0 & -1 & 0 & 0
   & 0 & 0 \\
 0 & 0 & 0 & 0 & 0 & 0 & 0 & 0 & 0 & -1 & 0 & 0 & 0
   & 0 & 0 \\
\end{array}
\right)\, .
\end{equation}

On the other hand the matrix $\Phi_{\mathrm{Fund}}(\Theta_{\mathrm{I}})$ acting on the fundamental representation reads
\begin{equation}
\Phi_{\mathrm{Fund}}(\Theta_{\mathrm{I}}) = \left(
\begin{array}{cccccccc}
 0 & 0 & 0 & 0 & 0 & 0 & 0 & 1 \\
 0 & 0 & 0 & 0 & 0 & 0 & 1 & 0 \\
 0 & 0 & 0 & 0 & 0 & 1 & 0 & 0 \\
 0 & 0 & 0 & 0 & 1 & 0 & 0 & 0 \\
 0 & 0 & 0 & 1 & 0 & 0 & 0 & 0 \\
 0 & 0 & 1 & 0 & 0 & 0 & 0 & 0 \\
 0 & 1 & 0 & 0 & 0 & 0 & 0 & 0 \\
 1 & 0 & 0 & 0 & 0 & 0 & 0 & 0 \\
\end{array}
\right) \, \ .
\end{equation}
In this case, the measure of the disconnected part, with the same parametrization of the fugacities \eqref{defz}, is
\begin{align}
 \mathrm{d} \mu_{4,\mathrm{I}}^{-}(z)&=\frac{\mathrm{d} z_1}{2\pi i z_1}\frac{\mathrm{d} z_2}{2\pi i z_2}\left(1+z_1^2\right)
   \left(1-z_2^2\right)
   \left(1-\frac{z_1^4}{z_2^2}\right)
   \left(1+\frac{z_2^2}{z_1^2}\right) \, .
\end{align}
Note the somewhat unusual $+$ signs that appear in the measure, as a consequence of (\ref{measure}).

\subsubsection{The Higgs branch Hilbert series of SQCD}

It is now straightforward to put all ingredients in place and explicitly evaluate \eqref{eq:Hb} to obtain the full, refined, Higgs branch Hilbert series (or Hall-Littlewood index). Due to the lengthy --and rather non-illuminating-- expressions, here we will quote some such examples of the type I projection for even $N$ cases, referring to \cite{Bourget:2019phe} for type I for odd $N$ and type II examples.
\paragraph{$N=4$ and $F=8,10$}
\begin{align*}
\textrm{HS}^{\mathrm{I}}_{(4,8)}(t;q_i)& =   1 + [2,0,0,0]_{C_4}t^2 + \Big([4,0,0,0]_{C_4}+2[0,2,0,0]_{C_4}+2[0,0,0,1]_{C_4}\\
& +3[0,1,0,0]_{C_4}+3\big)t^4 + o(t^4) \, \ , \\
\textrm{HS}^{\mathrm{I}}_{(4,10)}(t;q_i) & = 1 + [2,0,0,0,0]_{C_5}t^2 + \Big([4,0,0,0,0]_{C_5}+2[0,2,0,0,0]_{C_5}+2[0,0,0,1,0]_{C_5}\\
& +3[0,1,0,0,0]_{C_5}+3\Big)t^4 + o(t^4)\, \ ,
\end{align*}
where $\{q_i\}$ denote a set of global symmetry fugacities.
\paragraph{$N=6$ and $F=12,14$}
\begin{align*}
 \textrm{HS}^{\mathrm{I}}_{(6,12)}(t;q_i) &= 1 + [2,0,0,0,0,0]_{C_6}t^2 +\Big([4,0,0,0,0,0]_{C_6}+2[0,2,0,0,0,0]_{C_6}\\
& +[0,0,0,1,0,0]_{C_6}+2[0,1,0,0,0,0]_{C_6}+2\Big)t^4 +o(t^4)\, \ ,\\
 \textrm{HS}^{\mathrm{I}}_{(6,14)}(t;q_i) &= 1 + [2,0,0,0,0,0,0]_{C_7}t^2 + \Big([4,0,0,0,0,0,0]_{C_7} + 2[0,2,0,0,0,0,0]_{C_7} \\
& +[0,0,0,1,0,0,0]_{C_7} + 2[0,1,0,0,0,0,0]_{C_7} \Big)t^4 + o(t^4) \, \ . \\
\end{align*}

As can be seen in these expressions --as well as in the analogous ones in \cite{Bourget:2019phe}--, at order $t^2$ we find the character of the representation of the adjoint of the predicted global symmetry group. Since such contribution is precisely coming from the conserved global symmetry current multiplet (in fact from one of the scalars in the multiplet $\sim$ moment maps), and the latter must be in the adjoint by definition, this provides a further check on our expectations.

\subsubsection{The full unrefined Higgs branch Hilbert series}

Upon unrefining one can find a slightly more manageable form of the Higgs branch Hilbert series.

As for the component connected to the identity, which is identical to standard SQCD, the Highest Weight Generating function (HWG) is known exactly \cite{Hanany:2014dia,Bourget:2019phe}
\begin{equation}
\label{eq:hwgp}
\textrm{HWG}^{+}_{(N,F)}(t;\mu_i)=    \textrm{PE}\left[t^2+\sum_{i=1}^{N-1}t^{2i}\mu_i\mu_{F-i}+t^N(\mu_N+\mu_{F-N})\right]\, \ .
\end{equation}
Here the $\{\mu_i\}$ denote a set of highest weight fugacities for the $\mathrm{SU}(F)$ global symmetry group. Then, using (\ref{eq:hwgp}), we can obtain the expression of the corresponding Hilbert Series $\textrm{HS}^{+}_{(N,F)}(t)$ for the component connected with the identity. On the other hand the Hilbert Series for the component non-connected with the identity $\textrm{HS}^{(\mathrm{I},\mathrm{II}),-}_{(N,F)}(t)$ can be explicitly computed performing the integration with the corresponding measure.

As an explicit example, let us consider the case of $\widetilde{\mathrm{SU}}(3)_I$ with $F=6$, and, in order not to clutter the presentation, postpone to appendix \ref{app:Hs} a longer list of examples.

\paragraph{\textbf{$N=3$} and \textbf{$F=6$}}
\begin{align*}
    \textrm{HS}^{+}_{(3,6)}(t)&=\frac{1}{(1-t)^{20} (1+t)^{16} (1+t+t^2)^{10}}\Big(1 + 6 t + 41 t^2 + 206 t^3 + 900 t^4 + 3326 t^5 +  \\
    & 10846 t^6 +
 31100 t^7 + 79677 t^8 + 183232 t^9 + 381347 t^{10} + 720592 t^{11} +
 1242416 t^{12} +\\
 & 1959850 t^{13} + 2837034 t^{14} + 3774494 t^{15} +
 4624009 t^{16} + 5220406 t^{17} + 5435982 t^{18} +\\
 & ... + \ \textrm{palindrome}\  + ... + t^{36}\Big) \, \ , \\[15pt]
 \textrm{HS}^{\mathrm{I},-}_{(3,6)}(t) &= \frac{1+2 t^2+16 t^4+23 t^6+59 t^8+46 t^{10}+59 t^{12}+23 t^{14}+16 t^{16}+2 t^{18}+t^{20}}{(1-t^2)^{12} (1+t^2)^{8}}\, \ , \\[15pt]
 \textrm{HS}^{\mathrm{I}}_{(3,6)}(t) &= \frac{1}{(1-t)^{20} (1+t)^{16} (1+t^2)^{8} (1+t+t^2)^{10}}\Big(1 + 6 t + 34 t^2 + 144 t^3 + 647 t^4
 + 2588 t^5 +\\
 & 9663 t^6 +
 31988 t^7 + 97058 t^8 + 268350 t^9 + 687264 t^{10} + 1628374 t^{11} +
 3598201 t^{12} +\\
 & 7421198 t^{13} + 14364220 t^{14} + 26130494 t^{15} +
 44837750 t^{16} + 72656468 t^{17} + 111456702 t^{18} +\\
 & 162010222 t^{19} +
 223544610 t^{20} + 292994926 t^{21} + 365233973 t^{22} + 433158422 t^{23} + \\
 & 489154949 t^{24} + 526027956 t^{25} + 538960928 t^{26} + \ ... \  + \textrm{palindrome} \ + ... + t^{52}\Big)\, . \
\end{align*}

Note that again the coefficient of the $t^2$ term is just the expected one to reproduce the predicted global symmetry. Moreover, we observe that the dimension of the pole at $t=1$ is the same for both $\textrm{HS}^{+}_{(N,F)}(t)$ and $\textrm{HS}^{\mathrm{I}}_{(N,F)}(t)$. A similar feature, for a different type of disconnected group, was observed  in \cite{Bourget:2017tmt}. Moreover both the numerator of $\textrm{HS}^{+}_{(N,F)}(t)$ and the numerator of $\textrm{HS}^{\mathrm{I},-}_{(N,F)}(t)$ are given by a palindromic polynomial.

\section{Conclusions}\label{conclusions}

Because of a number of reasons, ranging from Condensed Matter inspirations to SUSY QFT, there has recently been interest in gauging discrete symmetries in Quantum Field Theory. In this paper we have discussed the case of the charge conjugation symmetry in gauge theories based on $\mathrm{SU}(N)$ gauge groups in a systematic manner (systematically extending \cite{Schwarz:1982ec} and its more recent stringy version \cite{Harvey:2007ab}). A key observation is that charge conjugation symmetry involves the outer automorphism of the $\mathrm{SU}(N)$ group, which is essentially complex conjugation and it is isomorphic to $\mathbb{Z}_2$. Since complex conjugation is non-trivially intertwined with the standard gauge transformations, it turns out that the appropriate framework for that is to construct a larger group which, from the beginning, includes both standard gauge transformations as well as complex conjugation on equal footing. More precisely, these two actions form a semidirect product group which can be thought as an extension of the outer automorphism group by the connected component. In this case this amounts to the extension of $\mathbb{Z}_2$ by $\mathrm{SU}(N)$. Quite surprisingly, and to our knowledge unnoticed in the literature, it turns out that the possible such extensions are in one-to-one correspondence with the Cartan classification of symmetric spaces (in this case of type A). Thus, in the case at hand it turns out that there are exactly two such groups including a gauged version of charge conjugation. Mirroring the terminology for symmetric spaces, we have dubbed these $\widetilde{\mathrm{SU}}(N)_{\mathrm{I,II}}$ (recall that the $\widetilde{\mathrm{SU}}(N)_{\mathrm{II}}$ only exists for even $N$). This extends \cite{Bourget:2019phe}, which, in the newest terminology, only considered $\widetilde{\mathrm{SU}}(N)_{\mathrm{I}}$ for odd $N$ and $\widetilde{\mathrm{SU}}(N)_{\mathrm{II}}$ for even $N$.

In this paper we provide an explicit construction of the $\widetilde{\mathrm{SU}}(N)_{\mathrm{I,II}}$ groups. As a by-product, we can explicitly write down the transformation properties of the fundamental and adjoint representations. Since these are the building blocks for 4d $\mathcal{N}=2$ SQCD-like theories, we can explicitly write down the Lagrangian and understand, from first principles, the global symmetry pattern. We find that the global symmetry for a $\widetilde{\mathrm{SU}}(N)_{\mathrm{I}}$ theory with $F$ (half)-hypermultiplets is $\mathrm{Sp}(\frac{F}{2})$ --which requires $F$ to be even--, while for a $\widetilde{\mathrm{SU}}(N)_{\mathrm{II}}$ theory with $F$ (half)-hypermultiplets it is $\mathrm{SO}(F)$. Also, the precise description of the groups allows us to write down a Haar measure and ultimately to explicitly compute indices counting operators which characterize some branches of the moduli space of the theory. Indeed, this way we can not only check that the expected global symmetry pattern emerges; but also that both $\widetilde{\mathrm{SU}}(N)_{\mathrm{I,II}}$ SQCD theories have non-freely generated Coulomb branches. This is very interesting as it provides examples of non-freely generated $\mathcal{N}=2$ Coulomb branches.

In this paper we focused, as a proof-of-concept, on 4d $\mathcal{N}=2$ SQCD-like theories based on $\widetilde{\mathrm{SU}}(N)_{\mathrm{I,II}}$. Nevertheless, it is clear that we are just scratching the tip of an iceberg. Staying in the, perhaps tamest, realm of $\mathcal{N}=2$ theories, it would be interesting to study the String/M-theory realization. The close relative of gauging the CP symmetry has been considered in string-theoretic constructions in the past (\cite{Strominger:1985it}. See \textit{e.g} \cite{Berasaluce-Gonzalez:2013sna} for a more recent discussion). While most of these constructions were typically devised with an eye on phenomenologically viable string-inspired scenarios, it is tempting to guess that our construction could fit along those lines.  Another natural embedding in String Theory is through an orientifold construction, where indeed our groups play a role at intermediary steps as discussed in \cite{Harvey:2007ab}. Yet another promising avenue would be embedded our theories into the class $\mathcal{S}$ framework, perhaps yielding a connection to the constructions in \cite{Zwiebel:2011wa,Mekareeya:2012tn}. It would also be very interesting to explore landmark aspects of discrete gauge theories such as codimension 2 defects \cite{Preskill:1990bm}, as well as other dimensionalities and other SUSY's (including no SUSY). In particular, in other dimensions it may be that new interesting phenomena are possible. For instance, given that $\pi_0(\widetilde{\mathrm{SU}}(N)_{\mathrm{I,II}})=\mathbb{Z}_2$, one may imagine a discrete $\theta$ parameter in a SUSY QM based on $\widetilde{\mathrm{SU}}(N)_{\mathrm{I,II}}$. Also, one may consider extending the construction to $\mathrm{U}(N)$ groups. Since the latter have a non-trivial fundamental group, the corresponding extended versions may lead to interesting phenomena. There may be also a parallel to the Pin groups, in particular upon considering quotients by subgroups of the center. Also more exotic constructions, similar to the 2d $\mathrm{O}(N)_{\pm}$ orbifolds as in \cite{Hori:2011pd}, may be possible in 2d. It would also be very interesting to explore dynamical aspects of these theories, perhaps using localization to compute correlation functions along the lines of \cite{Gerchkovitz:2016gxx}. We leave these avenues, and surely many more yet unnoticed, open to explore in further publications.

\section*{Acknowledgements}

We would like to thank C. Bachas, J.P. Deredinger and M.Lemos for useful discussions. G.A-T and D.R-G are partially supported by the Spanish government grant MINECO-16-FPA2015-63667-P. They also acknowledge support from the Principado de Asturias through the grant FC-GRUPIN-IDI/2018/000174. The work of A.B. is supported by STFC grant ST/P000762/1 and grant EP/K034456/1. A.P. is supported by the German Research Foundation (DFG) via the Emmy Noether program “Exact results in Gauge theories”.

\appendix

\section{Symmetric spaces and real forms}
\label{AppendixSymSpaces}

In this appendix we offer a lightning summary of some relevant facts on symmetric spaces. For a more thorough review, see Chapter 28 in \cite{bump2004lie}.

Let $G$ be a Lie group and $H$ a closed subgroup. In general, the quotient $G/H$ is not a group, but it is a well-behaved topological space, called a \emph{homogeneous space}. For instance, $\mathrm{SU}(N)/\mathrm{SU}(N-1)$ is the sphere $S^{2N-1}$ seen as the unit sphere of $\mathbb{C}^N$.\footnote{Similarly, $S^{N-1}$ seen as the unit sphere of $\mathbb{R}^N$ is $\mathrm{SO}(N)/\mathrm{SO}(N-1)$, and $S^{4N-1}$ seen as the unit sphere of $\mathbb{H}^N$ is $\mathrm{Sp}(N)/\mathrm{Sp}(N-1)$. }

Consider now the following situation: suppose $G$ is a connected Lie group, with an involution (i.e. an automorphism of order 2) $\Theta$ such that the subgroup $K=\{g \in G | \Theta (g) = g\}$ is compact. Then the homogeneous space $X=G/K$ is a \emph{symmetric space}, i.e. a Riemannian manifold in which around every point there is an isometry reversing the direction of every geodesic. The involution $\Theta$, and the corresponding involution on the Lie algebra $\mathfrak{g}$ of $G$, which we denote $\theta$, is called a \emph{Cartan involution}. Let $\mathfrak{k}$ be the Lie algebra of $K$, or equivalently the $+1$ eigenspace of $\theta$ in $\mathfrak{g}$. It is natural to also introduce the $-1$ eigenspace, that we call $\mathfrak{p}$. We have clearly
\begin{equation}
    \mathfrak{g} = \mathfrak{k} \oplus \mathfrak{p} \, .
\end{equation}
Now let's introduce another Lie algebra
\begin{equation}
    \mathfrak{g}_c = \mathfrak{k} \oplus i  \mathfrak{p} \, .
\end{equation}
Both $\mathfrak{g}$ and $\mathfrak{g}_c$ have the same complexification $\mathfrak{g}_{\mathbb{C}}$.
The involution $\theta$ induces an involution on $\mathfrak{g}_c$ defined by
\begin{equation}
\label{incolutiongc}
    x+iy \rightarrow x-iy
\end{equation}
where $x\in \mathfrak{k}$ and $y \in \mathfrak{p}$.

Now we go back to the level of the groups.
Under good assumptions, $\mathfrak{g}_c$ is the Lie algebra of a \emph{compact} and \emph{connected} Lie group $G_c$, and both $G$ and $G_c$ can be embedded in the complexification $G_{\mathbb{C}}$. Moreover (\ref{incolutiongc}) can be lifted to $G_c$, which means $X_c=G_c/K$ is also a symmetric space.

In summary, we have two symmetric spaces $X$ and $X_c$, one non-compact and one compact, which are said to be in \emph{duality} (see figure \ref{fig:sum}). For instance, the sphere $S^2$ can be realized as the compact symmetric space $\mathrm{SU}(2)/\mathrm{SO}(2)$, the hyperbolic plane $\mathcal{H}$ as the non-compact symmetric space $\mathrm{SL}(2,\mathbb{R})/\mathrm{SO}(2)$, and they are in duality. The duality between symmetric spaces is a generalization of this elementary example.

\begin{figure}[t]
\center{
\begin{tikzpicture}
    \node at (1, 0)   {Groups $\Theta$:};
    \node at (4, 0.5)    {non-compact space};
    \node at (4, 0)    {$K \subset G, \ X=G/K$};
    \node at (7,0) {$\xleftrightarrow{\text{duality}}$};
     \node at (10, 0.5)    {compact space};
    \node at (10, 0)   (b) {$K \subset G_{c}, \ X_{c}=G_{c}/K $};
    \node at (4,-1)      {$\Big\updownarrow$};
    \node at (10,-1)      {$\Big\updownarrow$};
     \node at (1, -2)   {Lie algebra:};
    \node at (4, -2)     {$\theta$ \ $\mathfrak{g}=\mathfrak{k} \oplus \mathfrak{p} $};
    \node at (7,-2) {$\xleftrightarrow{\text{same \ complexification}}$};
    \node at (10.5, -2)     {$\theta^{'}$ \ $\mathfrak{g}_c =\mathfrak{k} \oplus i\mathfrak{p}$};
\end{tikzpicture}
}
\caption{Summary of the duality relations between homogeneous space $X$ and $X_c$.\label{fig:sum}}
\end{figure}

The pairs of (irreducible, simply connected) symmetric spaces have been classified by Cartan. There are three types of pairs:
\begin{itemize}
    \item The Euclidean spaces;
    \item The pair with $G_c=(K \times K)/K$ and $G=(K_{\mathbb{C}})_{\mathbb{R}}$ where $K$ is a compact simple Lie group (a member of the Killing-Cartan ABCDEFG classification);
    \item A pair in Table 28.1 of \cite{bump2004lie}, which corresponds to the classification of noncompact real forms of the simple Lie algebras.
\end{itemize}
Here we are interested in those symmetric spaces where $G_c = \mathrm{SU}(N)$ for some $N$. Looking at the classification, we find that the candidates come from the third type, and are reported in the first three columns of Table \ref{tableSymSpaces}.

\section{Results for the unrefined Hilbert series}
\label{app:Hs}

In this appendix we collect the results obtained for the unrefined Hilbert series $\textrm{HS}^{\mathrm{I,II}}_{(N,F)}(t)$ with $N$ colors and $F$ flavors. Note that when $N$ is even both the type ${I}$-action and the type ${II}$-action are possible.

\paragraph{$N=3$,\ $F=8,10$  with action  $\Theta_{\mathrm{I}}$}
\begin{small}
\begin{align*}
& \textrm{HS}^{\mathrm{I}}_{(3,8)}(t)= \frac{1}{(1-t)^{32} (1+t)^{24} (1+t^2)^{12} (1+t+t^2)^{16}}\Big(1 + 8 t + 60 t^2 + 352 t^3 + 2180 t^4 + 12240 t^5\\
& + 63615 t^6 + 297072 t^7 + 1271655 t^8 + 5001104 t^9 + 18251874 t^{10} +
 62027176 t^{11} + 197358994 t^{12} +\\
 & 589894792 t^{13} + 1662662266 t^{14} +
 4431761456 t^{15} + 11202560833 t^{16} + 26916075192 t^{17} +
 61602528492 t^{18} +\\
 & 134547288976 t^{19} + 280922141406 t^{20} +
 561538929032 t^{21} + 1076105342885 t^{22} + 1979421972312 t^{23} +\\
 & 3498766636248 t^{24} + 5948607168296 t^{25} + 9737172113226 t^{26} +
 15357420491872 t^{27} + 23355546914320 t^{28} +\\
 & 34271353352936 t^{29} +
 48550884100169 t^{30} + 66437452982600 t^{31} + 87857610599498 t^{32} +\\
& 112323553804264 t^{33} + 138879963090049 t^{34} + 166117154759136 t^{35} +
 192266666483228 t^{36} +\\
 & 215374877940064 t^{37} + 233536846417860 t^{38} +
 245150314372704 t^{39} + 249146681474602 t^{40} \ ... \ + \\
 & \textrm{palindrome} \ + ...\  t^{80}\Big)\, \ ,\\[10pt]
& \textrm{HS}^{\mathrm{I}}_{(3,10)}(t)= \frac{1}{(1-t)^{44}(1+t)^{32} (1+t^2)^{16}(1+t+t^2)^{22}}\Big(1 + 10 t + 94 t^2 + 708 t^3 + 5594 t^4 + 40304 t^5 +\\
& 267596 t^6 +
 1604770 t^7 + 8823246 t^8 + 44685068 t^9 + 210162976 t^{10} +
 922138360 t^{11} + 3793387031 t^{12} +\\
 & 14685693384 t^{13} +  53699356234 t^{14} + 186024512912 t^{15} + 612212660929 t^{16} + 1918747129356 t^{17} +\\
 & 5739475779538 t^{18} + 16417980228736 t^{19} +
 44992209839201 t^{20} + 118311677930184 t^{21} +\\
 & 298973084347420 t^{22} +
 727001567961864 t^{23} + 1703229868953967 t^{24} +
 3848902875668712 t^{25} +\\
 & 8398044127896305 t^{26} +
 17709753210444906 t^{27} + 36126291437128415 t^{28} +
 71345154443802538 t^{29} +\\
 & 136509440283280531 t^{30} +
 253232898276985664 t^{31} + 455739121278331778 t^{32} +
 796177311646870288 t^{33} +\\
 & 1350951695000313907 t^{34} +
 2227550842846449570 t^{35} + 3570900466255197137 t^{36} +\\
& 5567741522682300884 t^{37} +
  8447064933353162776 t^{38} +
 12474366711895916734 t^{39} +\\
 & 17937609369569411305 t^{40} + 25123443718887660186 t^{41} + 34283553514238981759 t^{42} +\\
& 45592869670297954474 t^{43} +
 59103639171661870052 t^{44} +
 74701493375989226532 t^{45} + \\
 & 92071217634978085051 t^{46} +
  110680303143430918394 t^{47} + 129787178088343520066 t^{48} + \\
& 148478122575903878990 t^{49} +
  165732587105152093453 t^{50} + 180511607443936610316 t^{51} + \\   & 191859268605749303150 t^{52} +
 199003742403609087020 t^{53} + 201443245637522550224 t^{54} +\\
& + \ ... \ + \textrm{palindrome}+ \ ... + t^{108}\Big)\, \ .
\end{align*}
\end{small}

\paragraph{$N=4$, $F=8,10$  with action  $\Theta_{\mathrm{II}}$}

\begin{align*}
& \textrm{HS}^{\mathrm{II}}_{(4,8)}(t) = \frac{1}{(1-t^2)^{34}(1+t^2)^{17}}\Big(1 + 11 t^2 + 749 t^4 + 8520 t^6 + 123173 t^8 + 975504 t^{10} +\\
&
 7079801 t^{12} + 37130520 t^{14} + 168290287 t^{16} + 606231681 t^{18} +
 1880386783 t^{20} + 4837617956 t^{22} +\\
 & 10783278743 t^{24} +
 20384258878 t^{26} + 33595129641 t^{28} + 47516178744 t^{30} +
 58828027690 t^{32} +\\
 & 62834962052 t^{34}+ \ ... \ + \ \textrm{palindrome} \ + ... + t^{68}\Big)\, \ , \\[10pt]
& \textrm{HS}^{\mathrm{II}}_{(4,10)}(t) = \frac{1}{(1-t^2)^{50}(1+t^2)^{25}}\Big(1 + 20t^2 + 1880t^4 + 40559t^6 + 932570t^8 + 13749498 t^{10} + \\
& 172341355 t^{12} + 1684998864 t^{14} + 13851616125 t^{16} +
 94630953820 t^{18} +
  552972551103 t^{20} +\\
 & 2770203725095 t^{22} + 12073883443120 t^{24} + 45987359734926 t^{26} + 154444878746850 t^{28} + \\
 & 459222671967535 t^{30} + 1216126216507310 t^{32} +
 2877699662424911 t^{34} + 6109680294283385 t^{36} +\\
 & 11666292937742595 t^{38} + 20092424985476080 t^{40} +
 31261869088087670 t^{42} + 44025712808863775 t^{44} + \\
& 56169284503495746 t^{46} + 64994327796765700 t^{48} +
 68224551337259378 t^{50} + \ ... \ + \textrm{palindrome} \ + ... \ t^{100}\Big)
\end{align*}

\paragraph{$N=4$,  $F=8,10$   with action   $\Theta_{\mathrm{I}}$}
\begin{align*}
& \textrm{HS}^{\mathrm{I}}_{(4,8)}(t) = \frac{1}{(1-t^2)^{34}(1+t^2)^{17}}\Big(1 + 19 t^2 + 621 t^4 + 9672 t^6 + 115781 t^8 + 1012392 t^{10} +
 6929353 t^{12} +\\
&  37647616 t^{14} + 166763191 t^{16} + 610159441 t^{18} +
 1871499527 t^{20} + 4855440684 t^{22} + 10751422823 t^{24} + \\
& 20435224870 t^{26} + 33521903017 t^{28} + 47610887368 t^{30} +
 58717583354 t^{32} + 62951199956 t^{34}+\\
&  \ ... \ + \ \textrm{palindrome} \ + ... + t^{68}\Big)\, \
, \\[10pt]
& \textrm{HS}^{\mathrm{I}}_{(4,10)}(t) = \frac{1}{(1-t^2)^{50}(1+t^2)^{25}}\Big(1 + 30 t^2 + 1640 t^4 + 43719 t^6 + 903050 t^8 + 13965248 t^{10} + \\
& 171040855 t^{12} + 1691679084 t^{14} + 13821738043 t^{16} +
 94749067680 t^{18} + 552555331397 t^{20} +\\
& 2771531440035 t^{22} +  12070052718828 t^{24} + 45997431130604 t^{26} + 154420650803330 t^{28} + \\
 & 459276181907479 t^{30} + 1216017405986190 t^{32} +
 2877903862084869 t^{34} + 6109325929218841 t^{36} + \\
& 11666862552680995 t^{38} + 20091575715527008 t^{40} +
 31263044887405650 t^{42} + 44024199831283511 t^{44} + \\
& 56171095173235402 t^{46} + 64992311468943920 t^{48} +
 68226641217885546 t^{50} + \\
 &  + ... + \textrm{palindrome}   + ...   t^{100}\Big)
 \end{align*}

\paragraph{$N=5,  F=10$   with action   $\Theta_{\mathrm{I}}$}
\begin{small}
\begin{align*}
& \textrm{HS}^{\mathrm{I}}_{(5,10)}(t) = \frac{1}{(1-t)^{52}(1+t)^{48} (1+t^2)^{24}(1+t^2+t^3+t^4)^{26}}\Big(1 + 22 t + 284 t^2 + 2706 t^3 + 21955 t^4 +\\
& 160914 t^5 +
 1095989 t^6 + 6979246 t^7 + 41658165 t^8 + 233574566 t^9 +
 1234569365 t^{10} + 6174964900 t^{11} +\\
 & 29339025390 t^{12} +
 132880692724 t^{13} + 575483327555 t^{14} + 2389678052368 t^{15} +
 9537108858707 t^{16} +\\
 & 36658340475690 t^{17} + 135959694126589 t^{18} +
 487352408392372 t^{19} + 1690878189035940 t^{20} + \\
 & 5685865819978940 t^{21} + 18553353915421956 t^{22} +
 58812746144565240 t^{23} + 181295374749401949 t^{24} + \\
 & 543973294401568114 t^{25} + 1590097569959523153 t^{26} +
 4531884343550335332 t^{27} + 12602966622005009583 t^{28} +\\
 & 34222732445449084068 t^{29} + 90801798406026318027 t^{30} +
 235550865278275435154 t^{31} +\\
 & 597781158693692309598 t^{32} +
 1484941206577385534578 t^{33} + 3612556855586202953706 t^{34} + \\
 & 8611425331844868499654 t^{35} + 20123123002882735041990 t^{36} +
 46117967367942045961984 t^{37} +\\
 & 103701230641717242512770 t^{38} +
 228882161202628401398674 t^{39} + 496044838564012314603553 t^{40} + \\
 & 1056013484156029947574972 t^{41} + 2209065184079799283904974 t^{42} +
 4542364802182471087464116 t^{43} +\\
 & 9183898349160013048150427 t^{44} +
 18263102474622174109283076 t^{45} + 35731344980304035652518168 t^{46} + \\
 & 68797198502279183054832396 t^{47} + 130392515255665999661450280 t^{48} +
 243334669278371355251281076 t^{49} + \\
 & 447227865448283414970636444 t^{50} +
 809705050788821331767991526 t^{51} +
 1444419199557525884710374569 t^{52} + \\
 & 2539330673373178708242199168 t^{53} +
 4400400061161378562047041542 t^{54} + \\
 & 7517882728502831968413954866 t^{55} +
 12665124362834156846827184294 t^{56} + \\
 & 21043140994302550778160376372 t^{57} +
 34488341247592291002683019204 t^{58} + \\
 & 55765427955534322478937405226 t^{59} +
 88972575754699507596936788405 t^{60} + \\
 & 140091005097562491907119219110 t^{61} +
 217715405474146926177559832432 t^{62} + \\
 & 334004367894475080619545506914 t^{63} +
 505889843814484679526388720852 t^{64} + \\
 & 756580164670751794484968322138 t^{65} +
 1117380660642869503684842943144 t^{66} + \\
 & 1629837707627349788223179891574 t^{67} +
 2348184609132241268753454614722 t^{68} + \\
 & 3342030098734627900076782048544 t^{69} +
 4699182325256166921736227036528 t^{70} + \\
 & 6528444132112829102068477963998 t^{71} +
 8962152087005380332380602472677 t^{72} + \\
 & 12158166797853174056382018906264 t^{73} +
 16300962310475160862617427342532 t^{74} + \\
 & 21601416397311748248950447669348 t^{75} +
 28294881210629359687423170146093 t^{76} + \\
 & 36637125430881682085950853304052 t^{77} +
 46897794251366524433817202742132 t^{78} + \\
 & 59351139575785230185385458458708 t^{79} +
 74263933013520921469754722695984 t^{80} + \\
 & 91880686508283605721275742630480 t^{81} +
 112406560907646621311656207138740 t^{82} + \\
 & 135988625360814413094029942796482 t^{83} +
 162696416973845686429064538780835 t^{84} + \\
 & 192503011763013195188281840233904 t^{85} +
 225268022347578830193191985999101 t^{86} + \\
 & 260724052662385407911626535090360 t^{87} +
 298468136817513466381809076622798 t^{88} + \\
 & 337959547790685641993088340369146 t^{89} +
 378525073621853250977077476172104 t^{90} + \\
 & 419372430837618163454454120590622 t^{91} +
 459611939962498281014889114186630 t^{92} + \\
 & 498285964998009472402641743538914 t^{93} +
 534404969678972657531726866256818 t^{94} + \\
 & 566988428691645677376257138792706 t^{95} +
 595108314685264729771554857585174 t^{96} + \\
 & 617932519787766416628401096312304 t^{97} +
 634765409395059823386687683065751 t^{98} + \\
 & 645082773998319336209845681850552 t^{99} +
 648558747011165681457601756617802 t^{100} \ ... \ +\\
 & \textrm{palindrome} \ + ...  + t^{200} \Big)
\end{align*}
\end{small}

\paragraph{$N=6$, $F=12,14$  with action $\Theta_{\mathrm{I}}$}

We report the results only for the disconnected component $\textrm{HS}^{\mathrm{I},-}_{(N,F)}(t)$
\begin{align*}
& \textrm{HS}^{\mathrm{I},-}_{(6,12)}(t) =
\frac{1}{\left(1-t^2\right)^{42} \left(1+t^2\right)^{37}}\Big(1 + 7 t^2 + 69 t^4 + 358 t^6 + 2038 t^8 + 8419 t^{10} + 35209 t^{12}  \\
& +118646 t^{14} + 392133 t^{16} + 1091925 t^{18} + 2941220 t^{20} +
 6833264 t^{22} + 15255425 t^{24} \\
& +29803863 t^{26} + 55760142 t^{28} +
 92180215 t^{30} + 145662506 t^{32} + 204720814 t^{34} + 274750067 t^{36}\\
& + 329305773 t^{38} + 376711462 t^{40} + 385626520 t^{42} + \ ... \ \textrm{palindrome} \ ... \ + \ t^{84} \Big)\, \ ,
\\[10pt]
& \textrm{HS}^{\mathrm{I},-}_{(6,14)}(t) = \frac{1}{\left(1-t^2\right)^{54} \left(1+t^2\right)^{49}}\Big(1 + 9 t^2 + 101 t^4 + 654 t^6 + 4357 t^8 + 22320 t^{10} + 111704 t^{12}  \\
& + 469641 t^{14} + 1895000 t^{16} + 6669349 t^{18} + 22380498 t^{20} +
 66872433 t^{22} + 190076679 t^{24}\\
 & + 487466405 t^{26} + 1188492526 t^{28} +
 2638404185 t^{30} + 5568826504 t^{32} + 10772076177 t^{34}\\
 & + 19818706650 t^{36} + 33573603786 t^{38} + 54119513030 t^{40} +
 80595879849 t^{42} + 114256971885 t^{44}\\
 & + 149990270920 t^{46} +  187496330812 t^{48} + 217354673235 t^{50} + 239983501133 t^{52} + \\
 & 245894331898 t^{54} + \ ... \textrm{palindrome} ... \ + t^{108} \Big)\, \ .
\end{align*}

\bibliographystyle{unsrt}

\end{document}